\documentclass[twocolumn]{aastex631}

\usepackage{subfigure}
\usepackage{multirow}
\usepackage[11pt]{moresize}
\definecolor{light-gray}{gray}{0.80}

\newcommand{\hst}{HST}

\newcommand{\spitz}{Spitzer}
\usepackage{ascii}
\newcommand{\desk}{{\asciifamily DESK}}
\newcommand{\kms}{km\,s$^{-1}$}
\newcommand{\degrees}{$^{\circ}$}
\newcommand{\nii}{[N\,{\footnotesize II}]}
\newcommand{\oiii}{[O\,{\footnotesize III}]}
\newcommand{\fluxu}{erg s$^{-1}$ cm$^{-2}$}

\newcommand{\mum}{$\mu$m}

\received{12 8 2023}
\revised{14 11 2023}
\accepted{4 12 2023}

\shorttitle{A Multiwavelength study of HM Sge}
\shortauthors{Goldman et al.}

\begin{document}

\title{A Multiwavelength Study of the Symbiotic Mira HM Sge with SOFIA \& HST}

\correspondingauthor{Steven Goldman}
\email{sgoldman@stsci.edu}

\author[0000-0002-8937-3844]{Steven R. Goldman}
\affiliation{Space Telescope Science Institute, 3700 San Martin Drive, Baltimore, MD 21218, USA}
\affiliation{SOFIA-USRA, NASA Ames Research Center, MS 232-12, Moffett Field, CA 94035, USA}
\author[0000-0001-8858-1943]{Ravi Sankrit}
\affiliation{Space Telescope Science Institute, 3700 San Martin Drive, Baltimore, MD 21218, USA}
\author[0000-0003-2553-4474]{Edward Montiel}
\affiliation{SOFIA-USRA, NASA Ames Research Center, MS 232-12, Moffett Field, CA 94035, USA}
\author[0000-0002-2748-9809]{Sean Garner}
\affiliation{SOFIA-USRA, NASA Ames Research Center, MS 232-12, Moffett Field, CA 94035, USA}
\author[0000-0001-5201-3223]{Nathan Wolthuis}
\affiliation{SOFIA-USRA, NASA Ames Research Center, MS 232-12, Moffett Field, CA 94035, USA}
\author[0000-0003-3682-854X]{Nicole Karnath}
\affiliation{SOFIA-USRA, NASA Ames Research Center, MS 232-12, Moffett Field, CA 94035, USA}
\affiliation{Space Science Institute, 4765 Walnut St, Suite B Boulder, CO 80301, USA}
\affiliation{Center for Astrophysics Harvard \& Smithsonian, Cambridge, MA 02138, USA}

\begin{abstract}
We have targeted the dusty symbiotic Mira system HM Sge with four instruments from the IR to the UV. We have used these observations along with archival observations to study how the system has been evolving after its 1975 nova-like outburst. We have detected rovibrational water emission in a symbiotic system for the first time using new EXES high-spectral-resolution infrared spectroscopy. The features, detected in emission, have velocities consistent with the systemic velocity but do not show any clear evidence of high-velocity outflows. Mid-infrared photometry and grism spectroscopy show that the oxygen-rich asymptotic giant branch dust and dust output have shown little to no change over the past 39\,years. In the optical/UV, we detect three main \nii\ nebular features that were detected 22 years ago. Two of these features show a small amount of movement corresponding to average outflows speeds of 38\,\kms\ and 78\,\kms\ since they were previously observed; some previously detected \nii\ features are no longer visible. New UV spectroscopy has shown that the nebular environment continues to steadily relax after the system's 1975 outburst. The data suggest, however, that the temperature of the hot component has increased from 200,000 K in 1989 to greater than 250,000 K now. Our new and archival observations suggest that the evolution of the system after its outburst is swift with little to no major changes after a period of a couple of years.
\end{abstract}

\section{Introduction} \label{sec:intro}
Symbiotic systems are interacting binary systems composed of a cool giant and a hot companion (e.g. white dwarf, neutron star). These stars were initially classified as \emph{spectroscopically peculiar} by Annie Jump Cannon, as they hosted a combination of spectral features found in both hot and cool stars \citep{Merrill1940}. Since that time, over 350 symbiotic systems (Galactic and extragalactic) have been confirmed \citep[][and references therein]{Belczynski2000,Merc2019}.

Symbiotic systems provide a unique opportunity to simultaneously study a range of poorly understood astrophysical processes. The two stars' close proximity results in the transfer of material through accretion from the cool companion to the hot companion. This material can also build up on the hotter companion resulting in large nova-like outbursts or ``slow'' nova.

In some symbiotic systems the cooler companion is highly evolved, undergoing significant dust production and mass loss. In these cases the outflows, dust properties, and dust content can be greatly influenced by the hotter companion. These are referred to as dusty (D)-type symbiotic systems as opposed to the more common \citep[$\sim$80\%;][]{Belczynski2000} stellar or S-type \citep{Webster1975}. We can use identified symbiotic systems to study how these dustiest evolved giants known as thermally pulsing asymptotic giant branch (TP-AGB) stars are affected by an extreme environment.

Highly evolved AGB stars in their thermally pulsing phase typically have outflow velocities between 10 and 20\,km\,s$^{-1}$ \citep{Engels2015,Goldman2017,Olofsson2022}\footnote{\vspace{-0.25cm} We will henceforth refer to the TP-AGB phase of evolution when discussing the properties of AGB stars. \vspace{-0.5cm}}. Far faster winds have been detected in symbiotic systems, AGB stars transitioning into planetary nebulae, and the stages following the AGB \citep{Solf1983,Liimets2018,Guerrero2020,Sahai2022}. Episodic mass loss or recurrent novae can lead to winds with multiple components with velocities ranging from a few \kms\ to thousands of \kms \citep{Beliakina1978,Belyakina1979,Stauffer1984,Tomov2007,Tomov2017}.

Multiwavelength studies are critical for a comprehensive understanding of D-type symbiotics and their environments.  We describe here observations obtained under a joint Hubble Space Telescope (HST)/SOFIA Directors' Discretionary program targeting HM Sge. In this paper we will describe the unique symbiotic system HM Sge using new and archival multiwavelength data to better understand the system's local environment, kinematics, and dust. The paper is organized as follows: Section 2 outlines our new observations and archival datasets relevant to our discussion, Section 3 discusses our comparative analyses and new detections of IR water emission features, and Section 4 gives our concluding remarks. \\

\subsection{HM Sge}
HM Sagittae (Sge) is a nearby symbiotic system composed of a cool, highly evolved oxygen-rich (M-type) AGB star accreting material onto a white dwarf (WD). The system has a distance measured by Gaia Data Release 2 of $1027$\,pc \citep{Gaia2018}. A number of works have also attempted to determine the system's orientation and kinematics based on UV, optical, and radio data (Table \ref{table:motion}).

HM Sge is a symbiotic system with one of the most recent nova-like outbursts, occurring as recently as 1975 \citep{Dokuchaeva1976}. The system brightened six\,magnitudes in the optical, but unlike classical novae, the outburst remained near its peak brightness far longer than the expected few days \citep{Ciatti1979}. The outburst is suspected to be the result of a hydrogen shell flash on the accretion envelope of the WD \citep{Stauffer1984}.

Observations from the radio to X-ray have allowed us to study the environment around HM Sge as well as its individual hot and cool components \citep[see review by][]{Hinkle2013}. HM Sge's cooler AGB star component has a pulsation period of 527 days \citep{Munari1989, Murset1999}, an M7 spectral type, and is heavily reddened in the near-IR ($J-K$\,$\sim$\,3\,mag). These are all indicative of dust production, high mass loss, and a late stage of AGB evolution; less is known about the WD companion. \citet{Nussbaumer1990} noted a change in the effective temperature of the WD from 40,000\,K in 1976, just after the outburst, to 170,000\,K in 1989. They also found that the WD kept a consistent luminosity of $\ 10^4$\,$L_{\odot}$ during this time.

\begin{deluxetable}{lrr}
\tablewidth{\columnwidth}
\tabletypesize{\small}
\tablecolumns{3}
\tablecaption{Properties of HM Sge. \label{table:motion}}
\tablehead{
\colhead{Parameter} &
\colhead{Value} &
\colhead{Ref.}
}
\startdata
R.A. & 295.$^{\circ}$48783 & (1) \\
Decl. & 16.$^{\circ}$74436 & (1) \\ \vspace{0.25cm}
Distance & 1027$\pm$109\,pc & (1) \\
{\it Kinematics} \\
Proper motion & $-$0.443, $-$7.104 mas\,yr$^{-1}$ & (1) \\
Binary separation & 40 AU & (2)\rlap{\rm $^a$} \\
Binary axis orientation & 130\degrees\ E of N & (2)\rlap{\rm $^b$}\\
Binary axis rotation rate & ? & \\ \vspace{0.25cm}
Orbital period & $>90$\,yr & (3)\\
{\it AGB star} & \\
Pulsation period & 534\,days & (4)\\
Pulsation amplitude & $\Delta I$ $\sim$ 1 mag & (4)\\
Spectral type & M7 & (5)\\
Luminosity & 1500--2000 $L_{\odot}$ & (6)\\
Gas Mass Loss Rate & $4 \times 10^{-6}$\,${M}_{\odot}\,{\rm yr}^{-1}$ & (6) \\
\\
{\it White Dwarf} & \\
Temperature & $>$\,250,000\,K & (6) \\
\enddata
\tablenotetext{}{{\rm $^a$} Updated assuming the newer distance listed above. \\
{\rm $^b$} Assuming no rotation since the observations (1999).
\tablerefs{(1) \citet{Gaia2018},
(2) \citet{Eyres2001}, (3) \citet{Richards1999}, (4) \citet{Goldman_2022_research_note}, (5) \citet{Murset1999},
(6) \emph{This work}}\vspace{-0.9cm}}
\end{deluxetable}

\vspace{-0.25cm}\subsubsection{System Orientation}
The distance to HM Sge has been of great interest because the binary separation and speeds of outflows can be estimated based on angular displacements and the distance. While the dust and morphology of the system have hampered previous attempts to measure the distance, it has now been measured more accurately with Gaia with an estimated error of $\sim 10.6$\%\footnote{The morphology and large dusty convective envelope may result in underpredictions of the parallactic error. This tends to be more of a problem for far more-luminous AGB stars that have uncertainties greater than 20\% \citep{Andriantsaralaza2022}.}.

The binary orientation has also been a matter of debate. \citet{Solf1984} used narrowband UV spatial--spectral mapping to determined that the fastest outflows (post-outburst) were to the east and west of the system, suggesting a north/south binary axis. This was then corroborated by \citet{Richards1999} with extended radio continuum features also to the east and west and a model placing the Mira north of the WD. \citet{Schmid2000} then observed the polarization of the Raman O\,{\sc vi} line, which showed polarization at 33\degrees\ east of north\footnote{We will henceforth refer to all angles using this reference frame.} nearly parallel to the previously estimated binary axis, yet suggesting a binary axis perpendicular to it at 123\degrees. A similar binary angle was determined by \citet{Eyres2001}, who suggested that, based on the peak brightness in UV and optical \hst\ filters, the binary axis is around 130\degrees\ with the WD to the north at a separation of 40 mas (shown later in right panel of Figure \ref{fig:schematic}). Assuming the distance determined by Gaia, this yields an angular separation of 40\,AU. The binary axis shows a clear preference for a diagonal (NE or NW) orientation, something that has been seen in many observations of HM Sge. Evidence also suggests a value of the orbital period of at least 90\,yr \citep{Kenyon1986,Richards1999}. The geometry/morphology of the system is not well known, but evidence has suggested a torus-like structure seen either pole-on or edge-on \citep{Sacuto2007}.

\subsubsection{Outflows}
The winds measured around HM Sge have varied greatly since its outburst. A Wolf--Rayet--type wind was observed during the three years following the outburst with velocities up to 2000\,\kms\ \citep{Beliakina1978,Wallerstein1978,Belyakina1979,Stauffer1984}. \citet{Kwok_Leahy_1984} used X-ray fluxes from observations in 1979 to infer wind speeds based on estimates of the plasma temperatures of at least 700\,\kms. This was followed by a similar measurement and estimate for observations in 1992 that corresponded to wind speeds of greater than 500\,\kms\ \citep{Muerset1997}.

\begin{figure}
 \centering
 \includegraphics[width=\columnwidth]{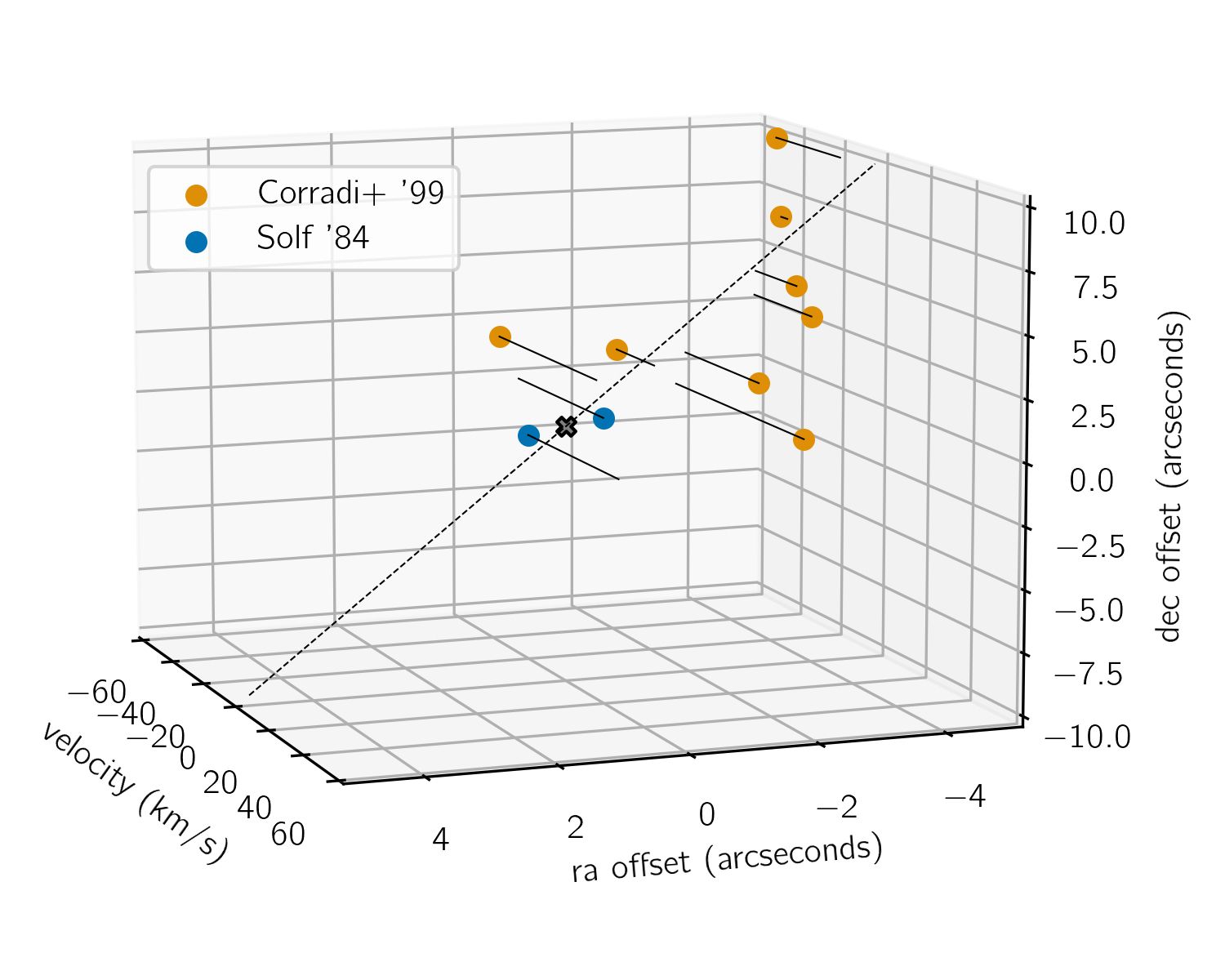}
 \caption{A subset of the velocities measured by previous observations of \nii\ emission. Emission is blueshifted above the dashed line and redshifted below, suggesting potential rotation and a binary axis parallel or perpendicular to this axis $\sim 120^{\circ}$.}
 \label{fig:3d_plot}
\end{figure}

In addition to the swiftly moving wind thought to be associated with the outburst are slower winds likely linked to current or previous mass loss. \citet{Cho2010} detected water and SiO maser emission with peak velocities of $-$94.4 and $-$98.9\,\kms, respectively. High-resolution spatial--spectral mapping in 1983 along directions corresponding to four slit angles was used to identify three different velocity components in \nii\ around 6548 and 6583\AA\ \citep{Solf1984}. Bipolar outflows of around 46\,\kms\ were observed to the east and west, linked to two narrow nebular lobes separated by 1.\arcsec5. Slower outflows of around 9.5\,\kms\ were also observed to the north and south, as well as a 2.6\,\kms\ outflowing component around the central region. Additional imaging and spectroscopy of \nii\ in 1996 show a range of velocities, with the most redshifted emission ($\sim 65$\,\kms) to the southwest and the most blueshifted ($\sim -50$\,\kms) to the northeast of the central nebula \citep{Corradi1999}. Some of these measured velocity components are shown in Figure \ref{fig:3d_plot}, and show a velocity profile consistent with a binary axis of around $120^{\circ}$ and a complex rotating morphology.

Raman-scattered O\,{\sc vi} lines (6825\AA) have also been detected in HM Sge \citep{Schmid2000,Lee2007}. \citet{Lee2007} found that these lines suggest a Keplerian disk around the WD star with an outer disk velocity of 26\,km\,s$^{-1}$ and a terminal velocity of the AGB component of 10\,\kms. In addition to plasma temperature and velocity-shifted emission, the proper motions of nebular features allow us to estimate average transverse velocities.

\begin{figure*}
 \centering
 \includegraphics[width=0.82\linewidth]{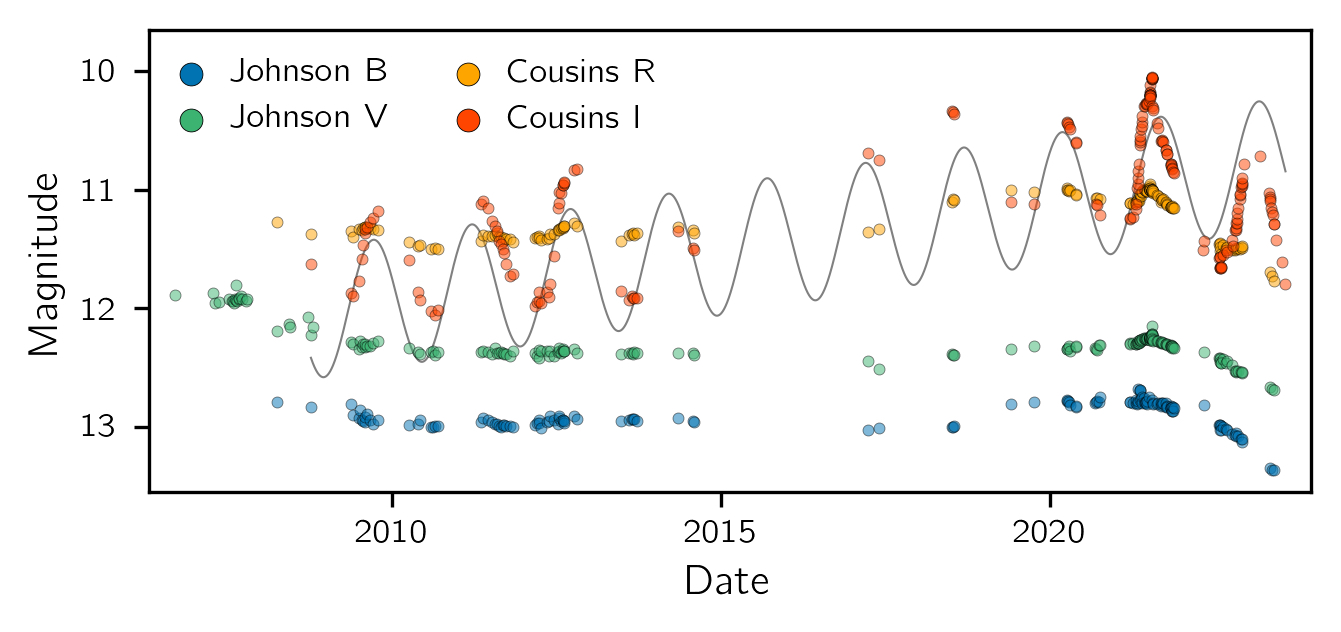}
 \caption{The brightness of HM Sge in {\it BVIR} filters from a subset of the available AAVSO data. We fit a sinusoidal function to the $I$ band data, showing a general increase in the IR flux independent of variability; the system, however, has dropped in brightness in the $I$-band. Adapted from \citet{Goldman_2022_research_note}. \\ }
 \label{fig:iband_trend}
\end{figure*}

\subsubsection{Nebular Features}
\citet{Kwok1984} and \citet{Richards1999} used the Very Large Array (VLA) to measure the motion of nebular features in radio continuum. \citet{Kwok1984} measured an average outflow of 0.\arcsec024\,yr$^{-1}$ away from the system in the NNW direction, or around $v_{\rm exp}\sim56$\,\kms\ ($\times D /$1\,${\rm kpc}$). \citet{Richards1999} used four epochs over five years to measure a movement of two discrete nebular features attributed to rotation, with an average value of $2.^{\circ}5\pm 0.^{\circ}5$ yr$^{-1}$. \citet{Richards1999} also detected features to the east and west that were brighter at low frequencies, suggesting that they may be nonthermal. They proposed that this may be synchrotron emission from the hot WD wind colliding with the dense blobs in the Mira wind.

HST/Faint Object Camera (FOC) optical images identified the presence of a slightly extended halo photoionized by the radiation from the hot component \citep{Hack1993}. HST/WFPC2 images obtained in 1999 showed nebular emission out to $\sim$2\arcsec\ with the brightest knots within 0.\arcsec 5 of the central binary system \citep{Eyres2001}. Ground-based emission line images in the \nii, \oiii, and H$\alpha$ filters were obtained using the Nordic Optical Telescope (NOT). The \nii\ images revealed the presence of a $\sim$30\arcsec\ nebula, indicating an earlier episode of mass loss from the central system. Spectra obtained at the same time showed that the kinematics in the inner nebula was consistent with a rotating inner torus whereas the outer nebula displays a string of knots indicative of episodic mass ejection \citep{Corradi1999}. Several works have also studied the conditions within the nebular environment. We will discuss these further in \S\ref{subsec:ultraviolet_emission}.

\subsubsection{Post-outburst Brightness}
HM Sge dimmed by several\,magnitudes in the optical after its outburst in 1975. In near-IR photometry from the AAVSO, however, the system increased in brightness by an average of 0.092\,mag yr$^{-1}$ from 2010 to 2022 (Figure \ref{fig:iband_trend}). The $I$-band light curve shows a fairly regular pulsation period. We fitted this light curve using a Lomb-Scargle technique to determine a pulsation period of $\sim 540$ days, similar to previous estimates \citep{Goldman_2022_research_note}. The $B$, $V$, $I$ and $R$ band photometry has also trended downward in the past several years. In the mid-1980s HM Sge also underwent a dimming in the infrared thought to be related to dust obscuration \citep{Munari1989}. This dimming may be linked to a recent mass loss event resulting in increased dust obscuration, or may possibly be linked to the orbital motion of the system.

\section{Observations} \label{sec:Observations}
We have observed HM Sge using the \hst and the Stratospheric Observatory for Infrared Astronomy \citep[SOFIA;][]{Temi2018}. Here we compare our new observations with archival observations taken over the last 50\,years. We compare new and old data from UV spectroscopy, near-UV\,+\,optical imaging, and IR photometry\,+\,spectroscopy, and present new high-resolution IR spectroscopy probing the outflow and accretion of water throughout the system.

\subsection{COS}
\label{subsec:COS}
Observations using the Cosmic Origins Spectrograph (COS) on \hst\ were obtained on 2021 March 21. Three positions were targeted -- one centered on the stellar system and the others on nearby nebular features to the north and south. Spectra were obtained with the low-resolution grating modes G140L/800 in the far-UV (FUV) and G230L/2950 in the near-UV (NUV). We did not detect any emission in the offset positions, and therefore in this paper we will restrict our attention to only the emission from HM Sge and the 2.\arcsec 5 diameter region around it that is included in the COS aperture.

The effective wavelength coverage of the FUV spectrum is 912--1948\AA\ obtained on a single detector segment. The NUV spectrum contains three stripes, with stripe A covering 1720--2125\AA\ and stripe B covering 2811--3214\AA. Stripe C is contaminated by second-order light, and did not provide useful data. The FUV and NUV data have spectral resolutions $\lambda/\Delta\lambda$ of about 1100 and 500, respectively. For our analysis, we use the calibrated data available from MAST, downloaded on 2022 October 7. The data have been reprocessed using the updated reference files that have been normalized to the latest CALSPEC models and have an improved extraction for the NUV data (COS STScI Analysis Newsletter, 2022 March). All of the \hst\ data used in this paper can be found in MAST:\dataset[10.17909/7gnq-a471]{https://archive.stsci.edu/doi/resolve/resolve.html?doi=10.17909/7gnq-a471}.

\subsection{WFC3/UVIS}
\label{subsec:WFC3/UVIS}
Observations using \hst's Wide-Field Camera 3 Ultraviolet--Visible channel (WFC3/UVIS) were taken on 2021 April 1. The observations included short exposures in three narrowband filters: F656N (H$\alpha$), F658N (\nii), and F502N (\oiii). These exposures are used to image HM Sge's inner nebula. An additional, longer exposure was taken in the F658N filter with the goal of reaching a similar sensitivity to the previous NOT observations. We use the calibrated data available from MAST, downloaded on 2021 June 9. These are used to study the morphology of the more extended surrounding nebula while saturating on the central source. We will compare these observations to archival observations from the NOT \citep{Corradi1999} and \hst /WFPC2 \citep{Eyres2001} to study changes in the nebular structure and measure proper motions of nebular features.

\subsection{FORCAST}
\label{subsec:FORCAST}
Observations were taken using the Faint Object Infrared Camera \citep[FORCAST;][]{Herter2018} on board SOFIA on the nights of 2021 July 1 and July 8. The observations include imaging in all ten available photometric filters, as well as grism spectroscopy with the FOR\_G111 (8.4--13.7\,$\mu$m) and FOR\_G227 (17.6--27.7\,$\mu$m) gratings.

\paragraph{Photometry} The imaging observations use standard chopping-and-nodding techniques to remove image artifacts and spurious detections. We perform aperture photometry for each of our filters using the standard reduction. Relative photometric statistical uncertainties were estimated using an annulus background (typically $\sim 5\%$). Photometric filters with their exposure times and effective wavelengths are listed in Table \ref{table:forcast} of the Appendix. The point-spread function estimated for the data is $\sim3.\arcsec 5$, and we find that the IR emission lies spatially within a similar radius.

\paragraph{Grism spectroscopy} Grism spectra were obtained in the FOR\_G111 and FOR\_G227 settings using the 4.\arcsec 7 wide slit, yielding spectral resolutions of $R = 110$ and $R = 130$, respectively. We use the standard calibrated Level 3 data provided by the SOFIA Science Center. Systematic errors for FORCAST grism spectroscopy can be as large as 10\% due to atmospheric conditions and slit misalignment. To get a more accurate absolute flux scaling we have normalized our spectra using the contemporaneous FORCAST photometry. We use the flux averages of the spectra, weighted by the filter transmission, and shift the spectra to the level of the FOR\_F111 and FOR\_F253 flux for the G111 and G227 spectra, respectively. This corresponds to shifts of 3.35 and $-$7.06 Jy for the G111 and G227 grisms, respectively. Our data are severely affected by atmospheric ozone between 9.2 and 10.3\,$\mu$m, and we exclude these data from our analysis.

\subsection{EXES}
\label{subsec:EXES}
Observations using the Echelon-Cross-Echelle Spectrograph \citep[EXES;][]{Richter2018} instrument on board SOFIA were taken on 2021 December 4, 2022 March 1, 2022 March 5, and 2022 April 27 with three unique and two duplicated settings. The angular resolution of the observations ranged from 1.\arcsec 8 to 2.\arcsec 3. All observations were taken with a channel width of $\sim$5\,nm and spanned a spectral range from 5.70 to 7.62\,$\mu$m with a resolution of $R$\,=\,60,000 ($v$ = 5.0\,\kms). The slit width was $\sim3.\arcsec 2$ and the slit length was between 8.\arcsec 85 and 9.\arcsec 6. We identify a number of rovibrational water emission features that probe the dusty outflows of the cooler component of HM Sge. We will use these to better understand the kinematics of the material surrounding HM Sge.

\subsection{Archival Observations}
\label{subsec:archival_observations}
Here we list the main observations used for our comparative analyses. For a more complete list of the archival observations of HM Sge see Table \ref{table:observations} of Appendix A.

\paragraph{NOT} The NOT was used for imaging in several narrowband filters (H$\alpha$, \nii, [O\,{\footnotesize I}], [O\,{\footnotesize II}], \oiii), as well as spectroscopy around the \nii\ $6583$\AA\ line at three positions and angles \citep{Corradi1999}. The data were not available online, but were provided by the authors (Romano Corradi, private communication). The data show clear detections of nebular features in the \nii\ filter with similar fainter structures in the [O\,{\footnotesize II}] filter. The spectra were used to measure radial velocities for the nebular features with velocities between $-$50 and 65\,\kms. Position-matching on bright sources was used to align the NOT images with our new WFC3 images.

\paragraph{ISO} The Infrared Space Observatory (ISO) was used to observe HM Sge in 1996 and 1997 \citep{Schild2001} with the Short Wavelength Spectrometer (SWS) and Long Wavelength Spectrometer (LWS), and the data show clear detections of the 10 and 18\,$\mu$m silicate features in the SWS data. We downloaded the SWS spectra from the ISO data archive; the longer-wavelength (43--196\,\mum) LWS data and \spitz\ MIPS data also exist, but are not discussed in this work as we have not obtained new longer-wavelength data for comparison. A number of nebular lines are detected in these spectra, including some from highly ionized species such as [Ne\,{\sc v}] and [Ne\,{\sc vi}]. The strongest line detected is [Ne {\sc vi}] $^2$P$_{3/2}-^2$P$_{1/2}$ at 7.6524\,$\mu$m, which was also observed with VLTI/MIDI \citep{Sacuto2007}.

\paragraph{IUE} The International Ultraviolet Explorer (IUE) was used to observe HM Sge many times between 1978 and 1992. The early data, analyzed by \citet{Mueller1985}, show that between 1979 June and September there was a dramatic increase in the brightness of the high-ionization lines He\,{\footnotesize II} 1640 \AA, C\,{\footnotesize IV} 1550 \AA, and N\,{\footnotesize V} 1240 \AA. The brightness of these lines increased rapidly up to 1985, and then remained at about the same level until 1989, while in the latter period, lower-ionization lines such as the semiforbidden N\,{\footnotesize III}] 1750 \AA, and O\,{\footnotesize III}] 1664 \AA\ lines (denoted with a single bracket) started declining in brightness \citep{Nussbaumer1990}. The model proposed by \citet{Nussbaumer1990} to explain the evolution of the UV emission line spectrum was photoionization of nebular material by the radiation from the hot star (and possibly an accretion disk) that was increasing in temperature while emitting at constant luminosity. \citet{Formiggini1995} presented a model for HM Sge that included both photoionization and a shock due to colliding winds expected in symbiotic systems \citep{Girard1987}. Their conclusions about the nature of the photoionization source agree with \citet{Nussbaumer1990}. Additionally they find that the lower-ionization line strengths are better explained if shock excitation is included, with the shock velocity decreasing from about 300~km~s$^{-1}$ to 180~km~s$^{-1}$ between 1979 and 1990. \citet{Murset1994} used the IUE data to estimate the temperature of the ionizing source at different epochs. They noted the presence of [Mg {\footnotesize VI}] lines in 1990 October, which were not there in 1989 November, and inferred that the radiation temperature had increased from 170,000 to 200,000\,K.

\begin{figure*}
 \centering
 \includegraphics[width=\linewidth]{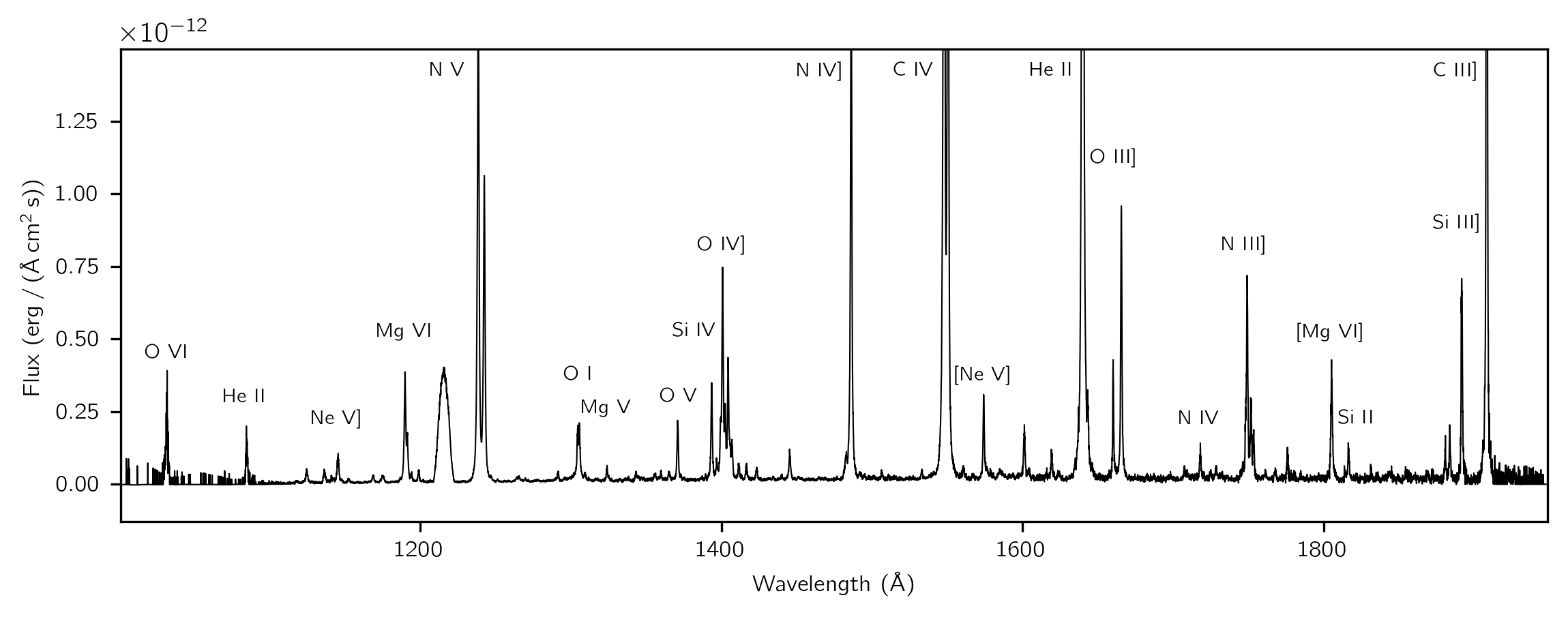}
 \includegraphics[width=\columnwidth]{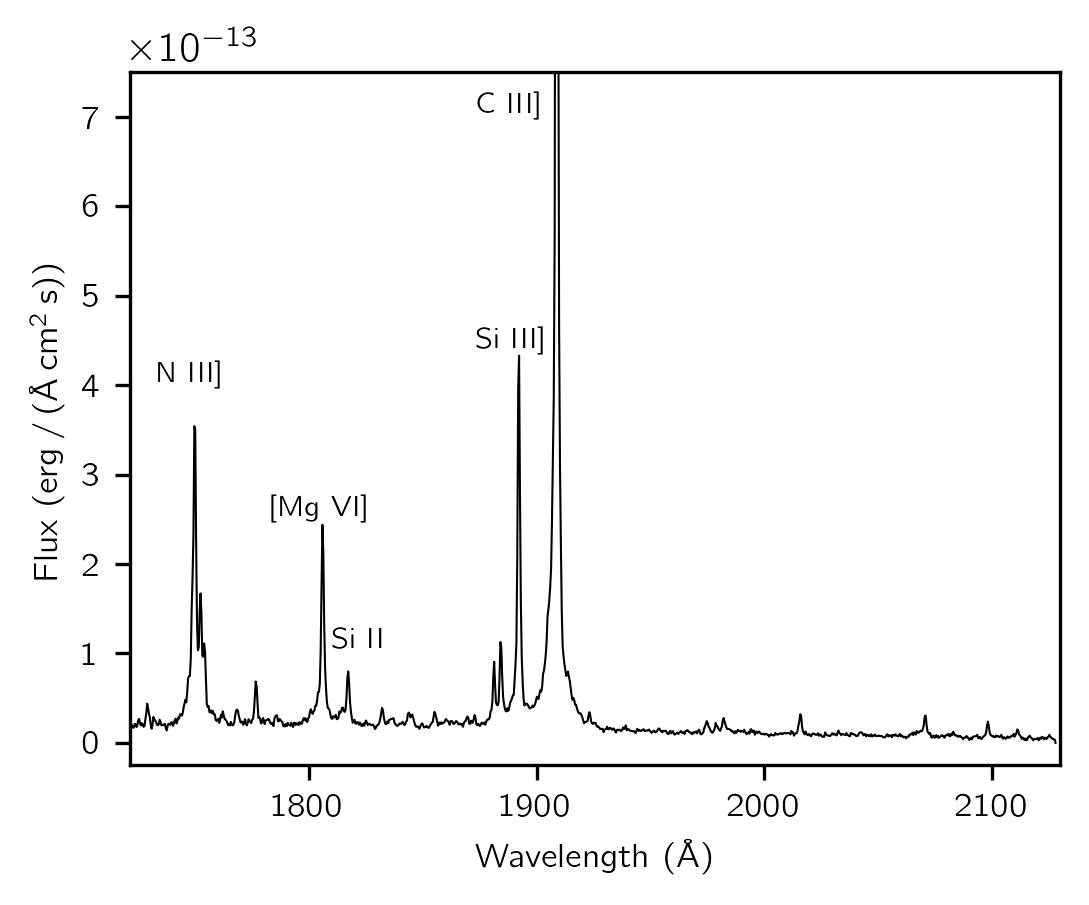}
 \includegraphics[width=\columnwidth]{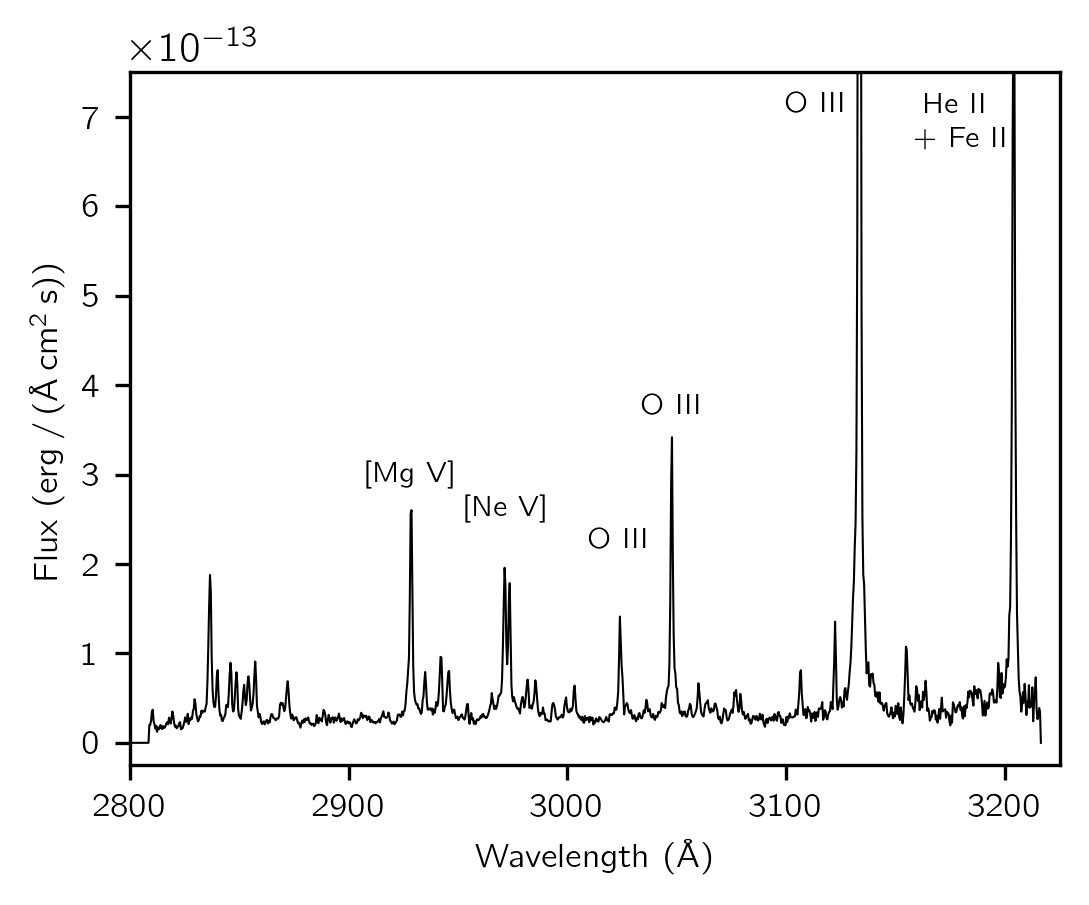}
 \caption{HST/COS NUV and FUV spectra showing the UV emission lines in the nebular environment around HM Sge. \\}
 \label{fig:uv_spectroscopy}
\end{figure*}


\begin{figure*}
    \centering
    \includegraphics[width=\linewidth]{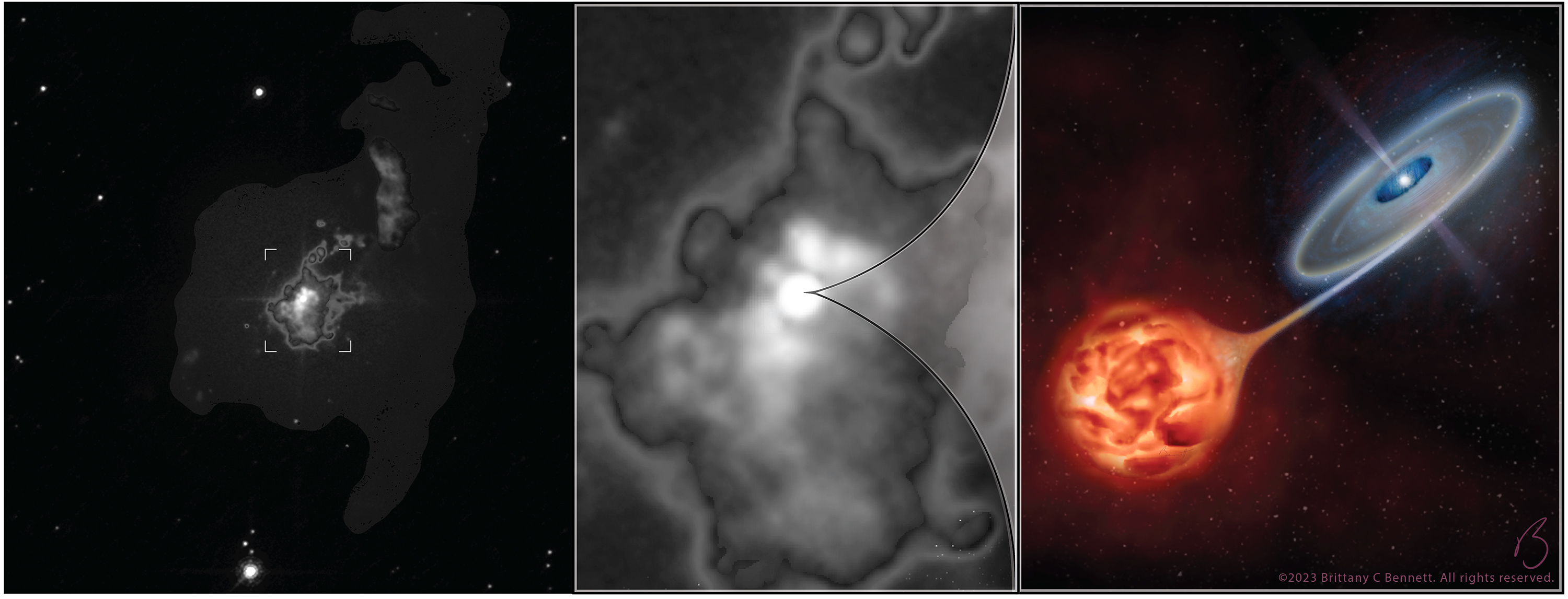}
    \caption{\hst\ WFC3 \nii\ image of HM Sge and the surrounding nebula enhanced artistically to accentuate the main nebular features of different contrast scales for the large-scale nebular structure (left) and central region (center). Also shown is an illustration of the inner stellar components of HM Sge (right), showing the suspected orientation and location of the hot and cold components; there are uncertainties in the orientation, which are discussed further in \S3.2. \copyright\ 2023 Brittany C. Bennett. All rights reserved. Used with permission. \\}
    \label{fig:schematic}
\end{figure*}

\section{Results and Discussion} \label{sec:results}
We will start by discussing the results that focus broadly on the environment surrounding the system, we will then move inward to discuss the kinematics of the material moving within and away from the system, and then discuss the properties of the stellar components of the system, particularly the dust surrounding the AGB star.

\subsection{Ultraviolet Emission}\label{subsec:ultraviolet_emission}

The calibrated COS spectra are shown in Figure~\ref{fig:uv_spectroscopy}. The top panel shows the G140L/800 FUV data, and the bottom panels show stripes A and B of the G230L/2950 NUV data. A large number of lines are detected and their fluxes span a wide range. Most of the strong and medium-strength lines are labeled in the figure. A number of the weak lines are from either Fe\,{\footnotesize II} or Fe\,{\footnotesize III}, or else are unidentified. In this paper, we do not consider any of these weaker lines, nor any of the medium-strength Fe lines, of which there are a few. Rather, we focus on lines detected in IUE spectra, and analyzed by \citet{Mueller1985} and \citet{Nussbaumer1990}

The measured fluxes of a selected set of lines are given in Table~\ref{table:uv_spec}. The last column tabulates the change in the observed line strengths between the COS spectra and the most recent IUE spectra presented in \citet{Nussbaumer1990}. In many cases, although doublet and multiplet lines are resolved in our spectra we tabulate the total flux for ease of comparison with the earlier measurements and discuss individual component strengths where relevant. For the region of overlap, we report measurements obtained using the FUV spectrum. The NUV Stripe A spectrum was used to check for consistency.

The most obvious change between the 1989 and 2021 spectra is the significant decrease in most of the line strengths, ranging from $-$20\% for the semiforbidden N\,{\footnotesize IV}]\,$\lambda$1485 to $-$78\% for the forbidden [Ne\,{\footnotesize IV}]\,$\lambda$1601 line. A few line strengths have stayed about the same, and two have increased: N\,{\footnotesize V}\,$\lambda\lambda$1238, 1242 by 10\% and He\,{\footnotesize II}\,$\lambda$1640 by 36\%. Another notable difference is that Mg\,{\footnotesize VI}\,$\lambda$1806, which was not detected in 1989, is fairly strong in the COS spectrum. Additionally, several lines are detected at wavelengths shortward of 1200\,\AA, which were not covered by IUE. Of these, O\,{\footnotesize VI}\,$\lambda$1032 and He\,{\footnotesize II}\,$\lambda$1085 lie in a region of low throughput and are only weakly detected.

A detailed study of the evolution of the UV spectra since 1989 is ongoing and will be the subject of a future publication. Here we limit ourselves to a general interpretation of the changes observed. The increase in the strength of the high-ionization lines is evidence that the ionizing source has continued to become hotter. The [Ne\,{\footnotesize V}]$\lambda$1575/[Ne\,{\footnotesize IV}]$\lambda$1601 flux ratio suggests that the effective temperature now is at least 250,000~K (Table 5 of \citet{Nussbaumer1990}), whereas it was less than 200,000~K in 1989.

There are probably multiple reasons for the decrease observed in many emission line fluxes. Some of it may be due to the gas being ionized to higher states. This would explain, for instance, the decrease in both [Ne\,{\footnotesize V}] and [Ne\,{\footnotesize IV}] as a larger fraction of the gas would be ionized up to Ne\,{\footnotesize VI}. The separation between the two components may have increased due to orbital motion, which would imply a weaker photoionizing flux at the nebular interface and a decrease in line strengths. The COS data are not sufficient to discriminate between these scenarios or to address the possibility of changes in elemental abundances leading to changes in line strengths.

\citet{Formiggini1995} found that the wind collision shock had grown weaker with time since the explosion, but still contributed significantly to the low-ionization lines observed in the 1987--1988 IUE spectra. Specifically their models showed a decrease in the shock velocity from around 300~km~s$^{-1}$ in 1979 to 180~km~s$^{-1}$ a decade later. The continued presence of a shock, with even a still lower velocity, would be a possible reason why some of the lines have not decreased in strength because they would be collisionally excited rather than photoionized. This would help explain why the O\,{\footnotesize III}]$\lambda\lambda$1660, 1666 line strengths have remained constant while the O\,{\footnotesize III} Bowen fluorescence lines have become significantly fainter.

\subsection{Tracing Ionized Ejecta with \nii}
\begin{figure}%
\centering
\includegraphics[width=\columnwidth]{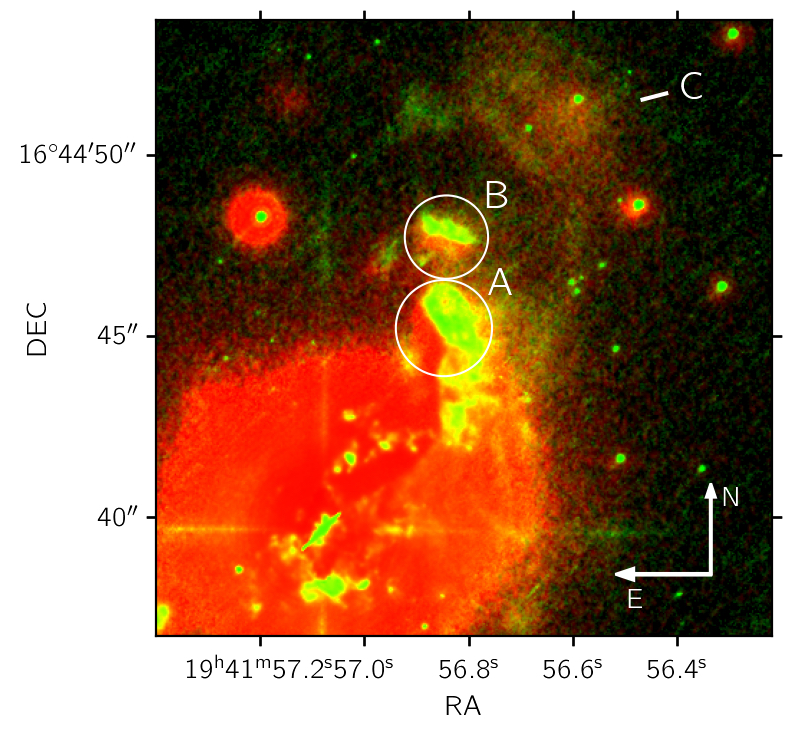}
\caption{A red (NOT)/green (HST/WFC3) image of the northwest region of HM Sge in narrowband \nii\ optical filters. The image shows the evolution of three main nebular features (that we identify as A, B, and C) over a period of 22\,years. The vertical/horizontal crosshair pattern around the central source (lower left) is the result of saturation.}
\label{fig:green_red}
\end{figure}

\begin{figure}
 \centering
 \includegraphics[width=\columnwidth]{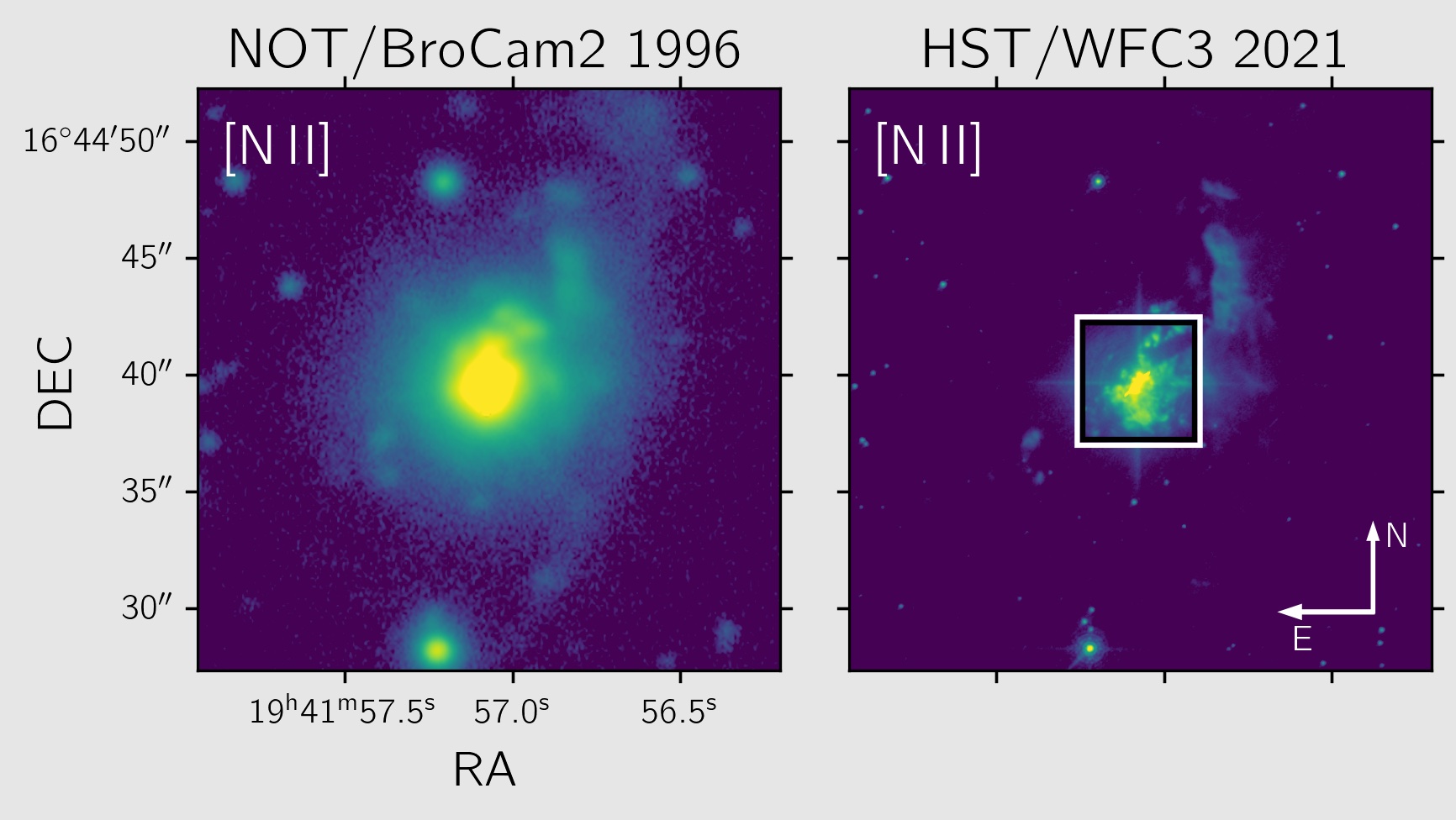}\\ \vspace{0.25cm}
 \includegraphics[width=\columnwidth]{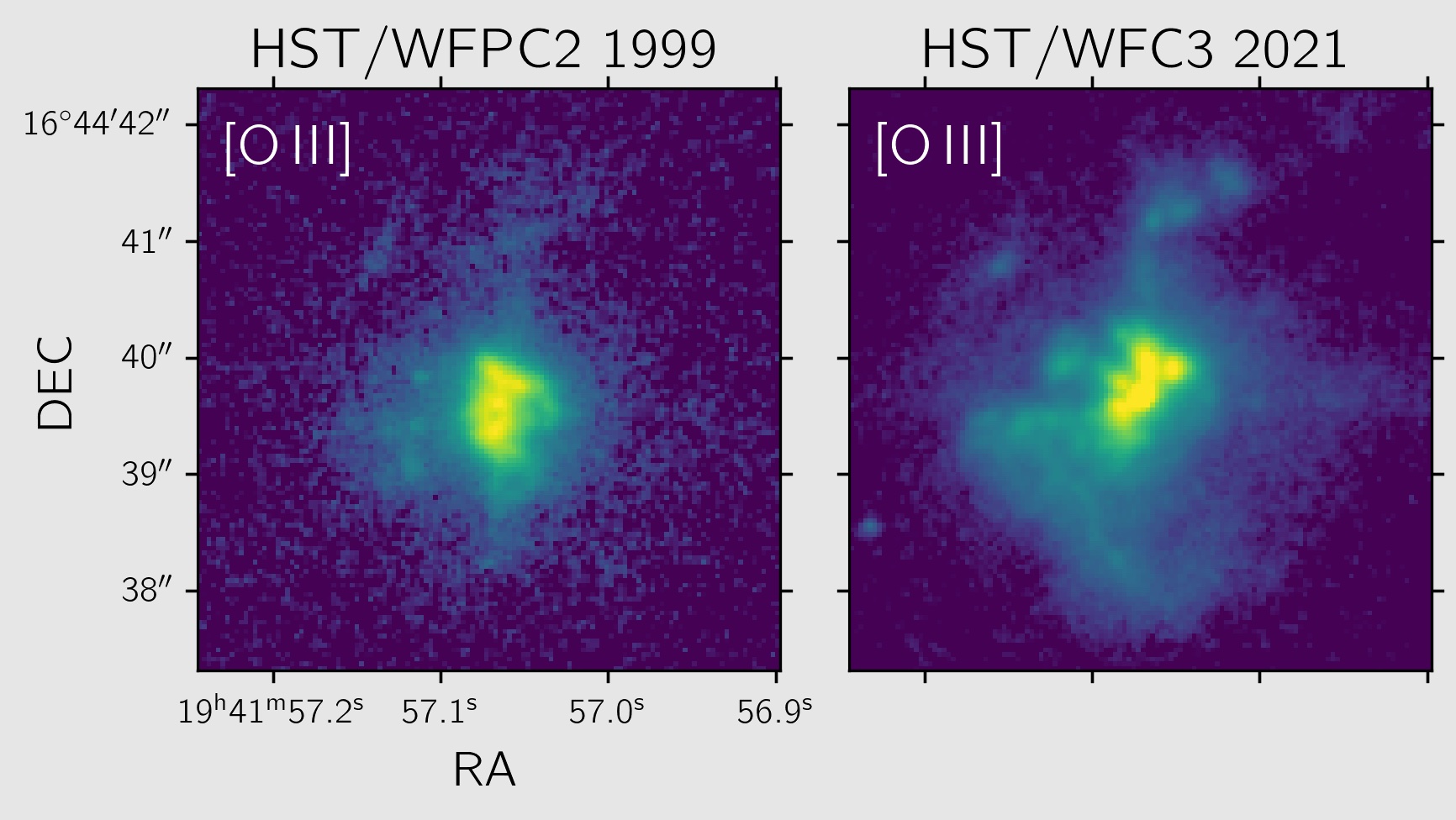}
 \caption{A comparison of new and archival narrowband optical images of HM Sge. The white box is the zoomed-in region shown in the lower images; intensities are displayed on a logarithmic scale.}
 \label{fig:wfpc2_comparision}
\end{figure}

Our new \hst\ WFC3 \nii\ imaging data of HM Sge show a variety of structures at differing levels of brightness  (Figure \ref{fig:schematic}). When we compare 1996 NOT observations to the 2021 WFC3 \hst\ observations, we can identify several prominent common nebular features to the northwest of the system (Figures \ref{fig:green_red} \& \ref{fig:wfpc2_comparision}); we will refer to these as features A, B, and C. These features show a small amount of movement between 1996 and 2021. Several additional smaller and faint common features are visible to the southeast and southwest.

\begin{figure}
 \centering
 \includegraphics[width=0.335\linewidth]{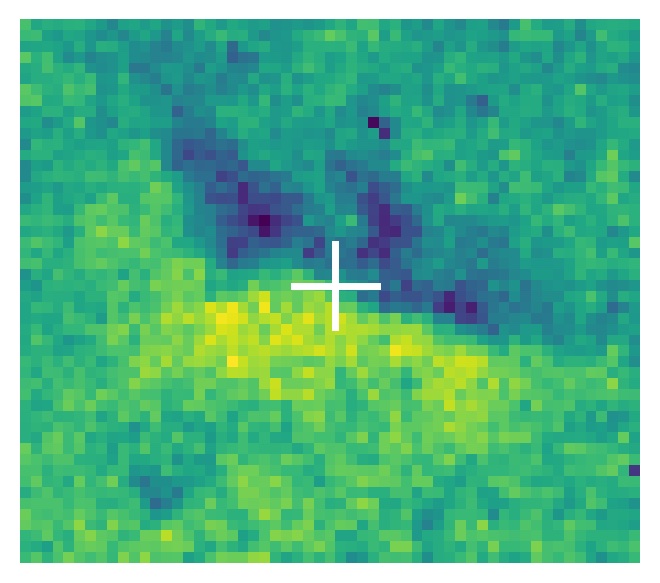} \hspace{-0.25cm} \vspace{-0.25cm}
 \includegraphics[width=0.335\linewidth]{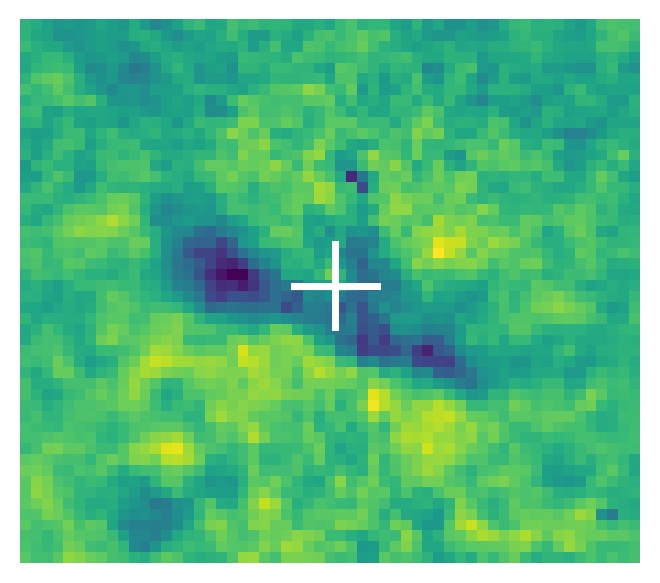} \hspace{-0.25cm}
 \includegraphics[width=0.335\linewidth]{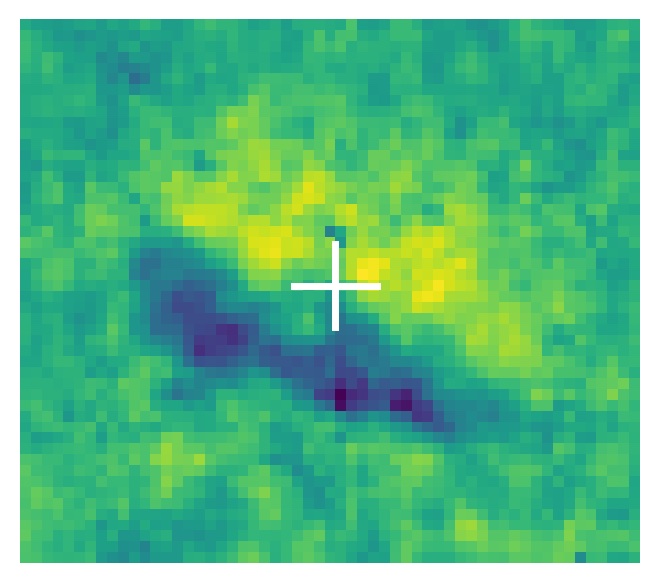} \\ \hspace{-0.25cm}
 \includegraphics[width=0.34\linewidth]{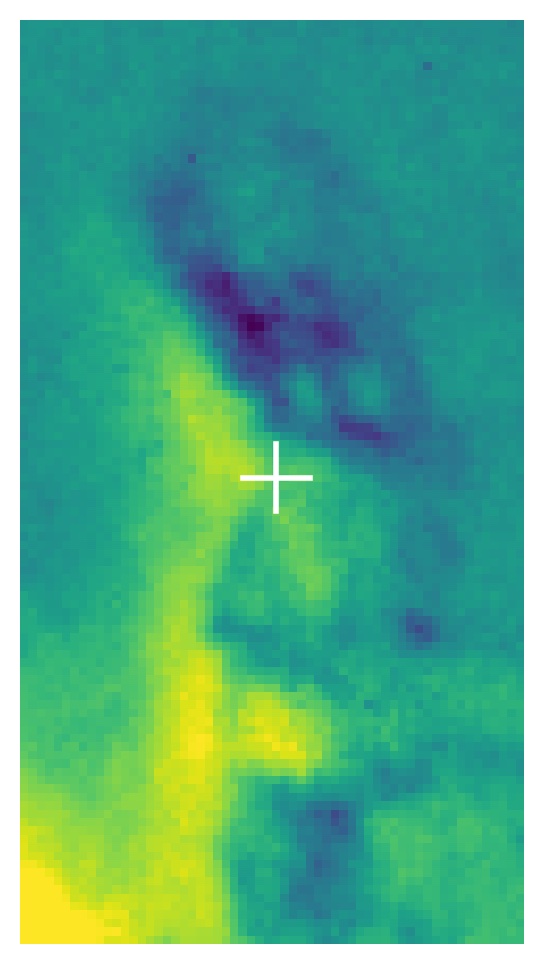} \hspace{-0.29cm}
 \includegraphics[width=0.34\linewidth]{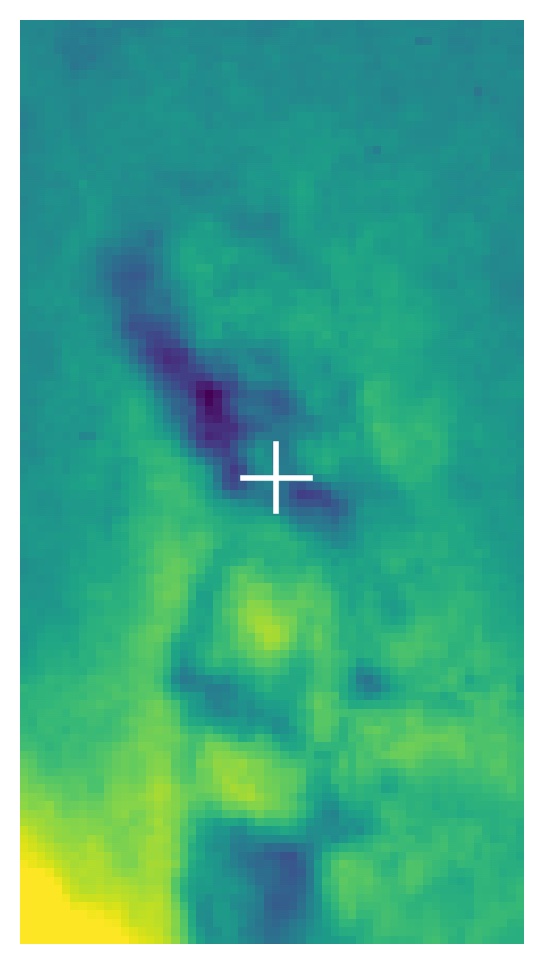} \hspace{-0.29cm}
 \includegraphics[width=0.34\linewidth]{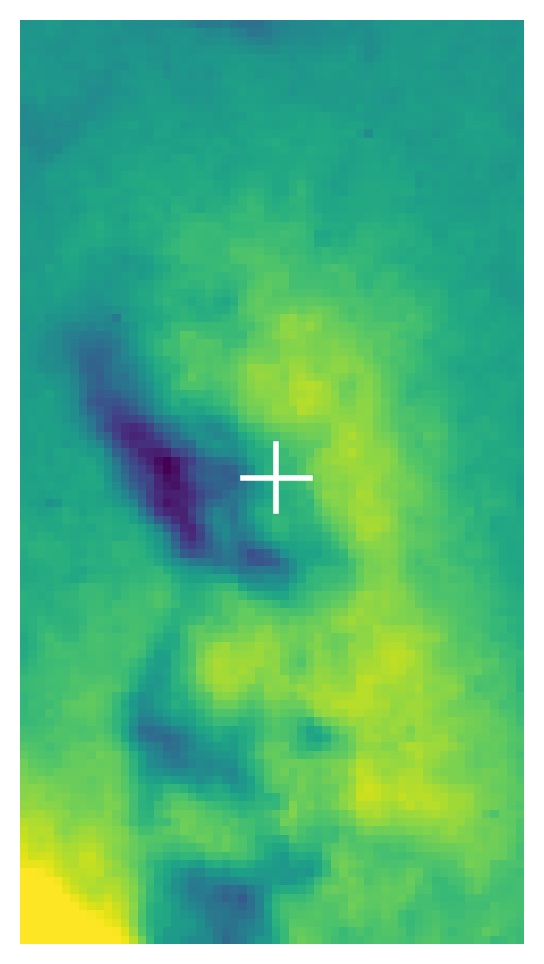} \hspace{-0.29cm}
 \caption{The (de)magnification technique used to fit the spatial shift of features A (bottom) and B (top) between the WFPC2 (1996) and WFC3 (2021) \nii\ images. The initial normalized image differences (left) are shown along with the best-fitted shift image (center) as well as a further demagnified source (right). Darker features are from WFC3 data that move down and to the left as we decrease the WFC3 pixel scale.\\}
 \label{fig:mag_fit}
\end{figure}

We have used a\,magnification technique similar to that of \citet{Reed1999} and \citet{Palen2002} to estimate the average outflow speed of features A and B. By taking a region of the WFC3 data, iteratively decreasing the pixel scale (demagnifying), degrading it to the resolution of the NOT, subtracting the image, and calculating the standard deviation, we are able to determine the best-fitted spatial shift of features A and B (Figure \ref{fig:mag_fit}). This technique allows for an estimate of the shift with precision below that of the NOT pixel scale. Using this technique we find best-fitted spatial shifts of 0.\arcsec40 and 0.\arcsec19 for features A and B, respectively. Using the time difference between the observations yields average outflow speeds for features A and B of 78\,\kms\ and 38\,\kms, respectively.

Given the distance to the system and the angular distance from the central region to the most prominent common features (A: 6.\arcsec3, B: 8.\arcsec7, C: 13\arcsec) these would have to have been moving at an average velocities of 1500, 2000, and 3000\,\kms\, respectively, since the outburst to be at the distances observed in 1996. Outflows at these speeds have not been detected since the outburst, suggesting they are from the 1975 outburst or, more likely, previous outbursts. The outflow speed of feature A, and the lower outflow speed for feature B farther out, are likely the result of shocks. Fast winds from previous outbursts have likely interacted with the material at large radii, slowed down and thermalized, and contributed to the slow outward movement and evolution of the nebula.

Data have shown evidence of $\sim$100\,\kms outflows possibly linked to jets. Outflows with these speeds have been detected in the radio \citep{Cho2010} and optical \citep{Corradi1999}, and inferred in the X-ray \citep{Toala2023}. We do not detect any jet-like outflows blowing out material at high speeds. While this does not discount the existence of jets, if they do exist, we do not see any evidence of them where we detect nebular emission. Higher-speed outflows may also be blowing out material in regions void of nebular material. Any small movement in the nebular features that we see may be attributed to the difference in resolution, ionization of the inner nebula, or slower movement of these features. We also have data in the H$\alpha$ and \oiii\ filters. The H$\alpha$ data are primarily saturated with one common feature north of the central nebula. The \oiii\ data also show several common features but no extended emission outside of 3.\arcsec5.

By comparing our F502N \oiii\ data with the 1999 WFPC2 observations we get a better view of the central region. We see a similar morphology to that seen in 1999 but seemingly rotated in the clockwise direction. This may suggest a rotation opposite to what was determined by \citet{Richards1999}, but may also reflect different parts of the nebula changing in brightness in \oiii. We also see additional extended emission that may be new or below the WFPC2 detection threshold. To better understand the outflows of the system, we have also targeted the rovibrational lines of water in the IR to measure the movement of material closer to the stars. \\

\begin{figure*}
 \centering
 \includegraphics[width=0.95\linewidth]{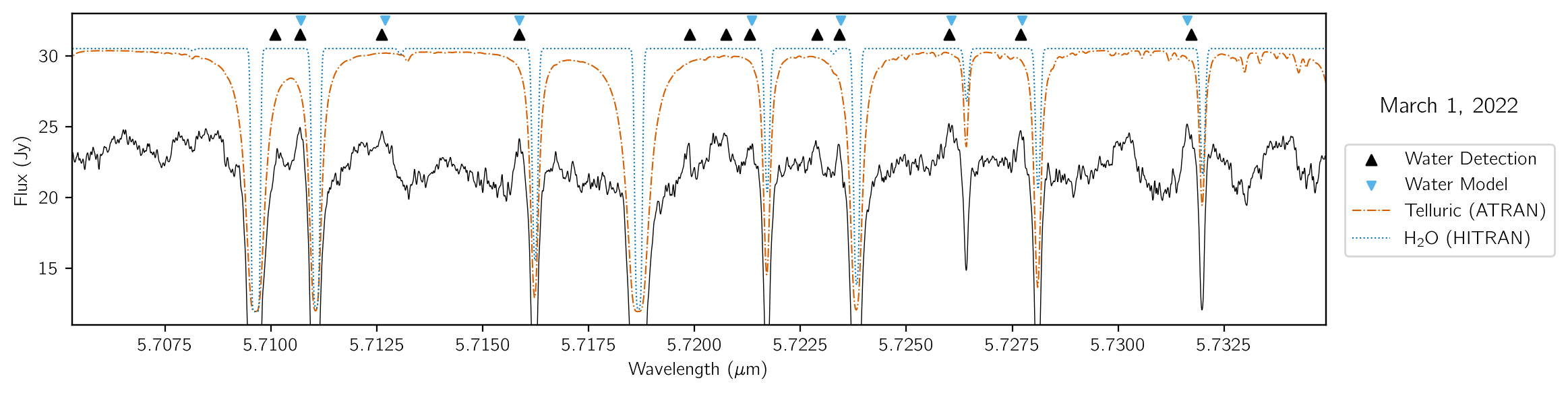}\\ \vspace{-0.2cm}
 \includegraphics[width=0.95\linewidth]{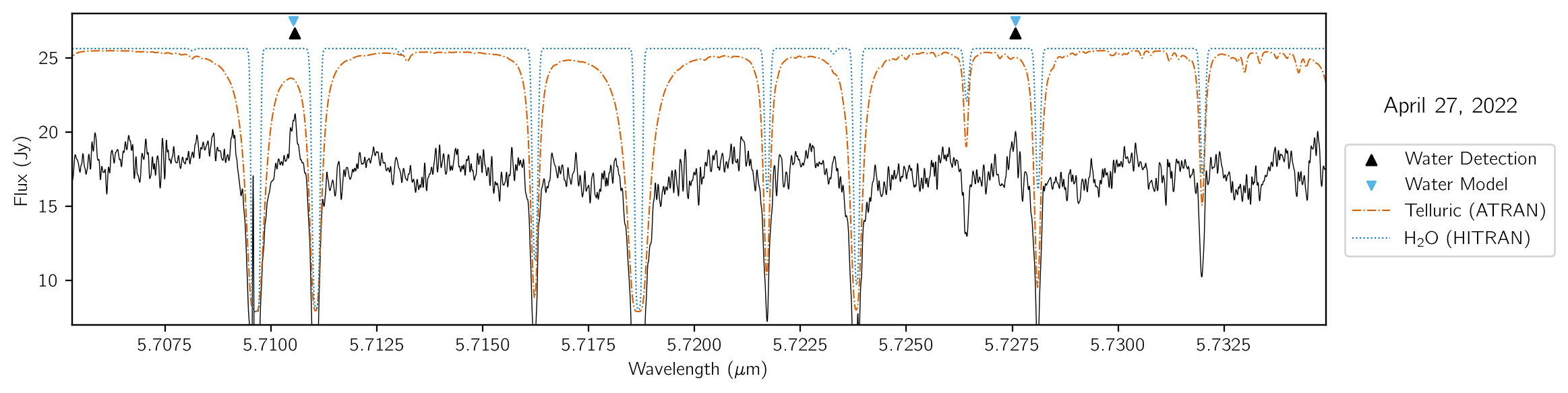}\\ \vspace{-0.2cm}
 \includegraphics[width=0.95\linewidth]{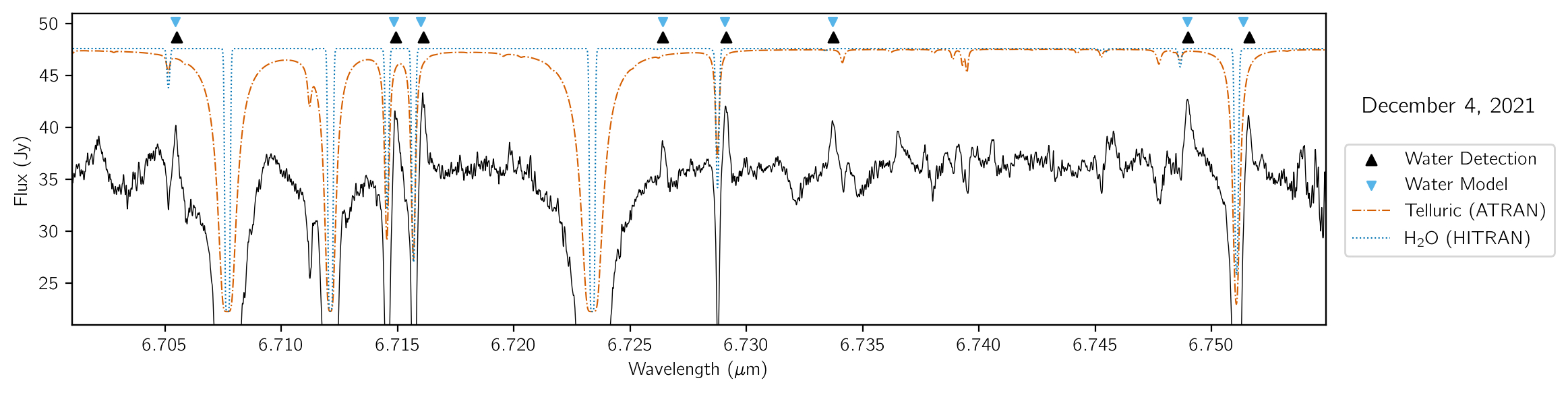}\\ \vspace{-0.2cm}
 \includegraphics[width=0.95\linewidth]{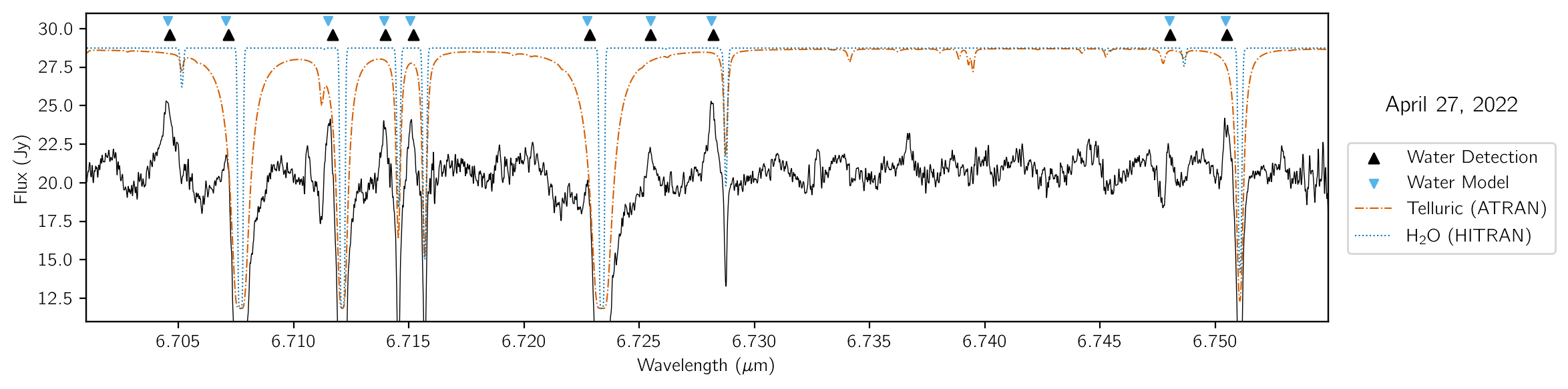}\\ \vspace{-0.2cm}
 \includegraphics[width=0.95\linewidth]{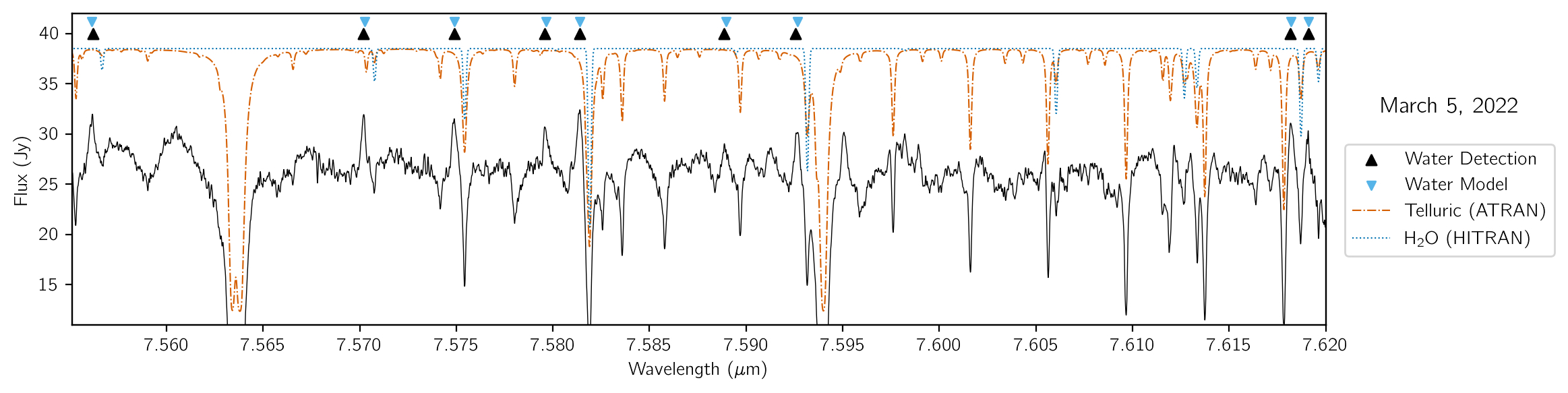}
 \caption{Infrared water emission features detected in high-resolution EXES spectra (black), which probe the kinematics of material around HM Sge. We show the ATRAN telluric model spectra (red) and HITRAN model spectra of water (blue). The features are velocity-shifted by the seasonal motion of the Earth, and the observations were scheduled so that these features fall in the wings of telluric features. The wavelengths of suspected water emission features and their corresponding model features at the system velocity are shown with black and blue triangles above the features.\\}
 \label{fig:exes_spectra}
\end{figure*}

\begin{figure}
 \centering
 \includegraphics[width=\linewidth]{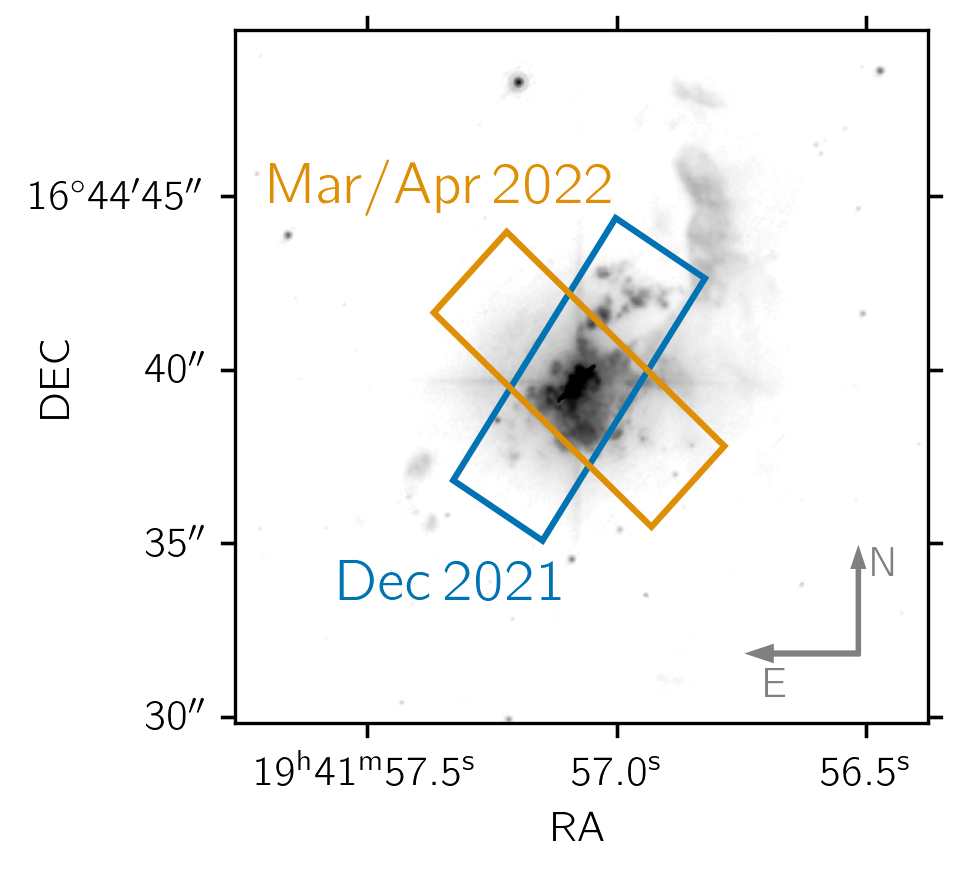}
 \caption{The EXES slit position for the two epochs of observation using the 6.7\,$\mu$m setting overlaid on our new long-exposure \hst\ \nii\ image.}
 \label{fig:exes_slit_position}
\end{figure}

\subsection{Wind Velocities from H$_2$O emission}
HM Sge was targeted with EXES to detect velocity-shifted rovibrational water lines that allow us to probe the kinematics of the winds surrounding the AGB and accretion disk. The four observations in three unique settings are displayed in Figure \ref{fig:exes_spectra}. In the figure we also show ATRAN telluric model spectra, as well as molecular transitions of water from HITRAN. We find that most of the emission features have small offsets ($< 10$\,\kms) from our assumed systemic velocity\footnote{The spectra are corrected for the velocity shifts from the seasonal motion of the Earth.} of $-5$\,\kms \citep{Hinkle2013}, giving more evidence for this as the systemic velocity. We also detect three lines with velocity shifts of $\sim 30$\,\kms; all of these lines are listed in Table \ref{table:exes}.

The small velocity shifts that we see in emission may be probing the material moving away from the cool component. In spherically symmetric AGB stars, we expect a slow dense wind expanding radially outward as the star loses mass and produces dust. This wind may be even slower along the line of sight for a source like HM Sge, which likely has higher-velocity bipolar outflows in the NW or NE direction. Assuming spherical symmetry and lines that are optically thick, these features should peak at the systemic velocity \citep{Wannier1990}. EXES data of other molecular features in nearby AGB stars \citep[e.g.][]{Fonfria2008} have shown emission features at or near the systemic velocity, and we see this in most of our water emission lines. The emission features have narrow peak widths representative of outflows with little to no outward velocity.

\begin{figure*}
 \centering
 \includegraphics[width=\linewidth]{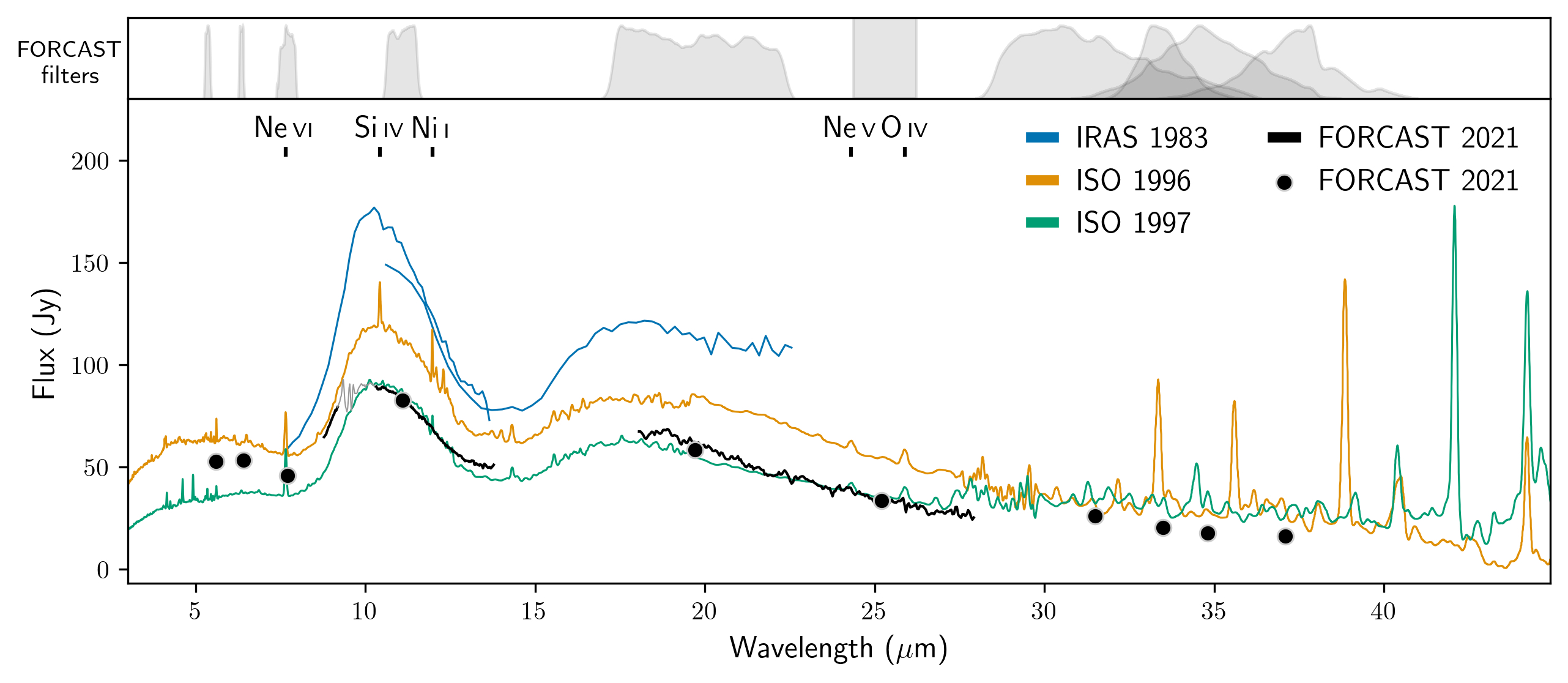}
 \caption{The mid-IR observations of HM Sge. The new FORCAST photometry (red circles) has statistical uncertainties of $\sim 5\%$. The spectroscopic observations include the region from 9.2--10.3\,$\mu$m, which is severely affected by atmospheric ozone and is ignored in our analysis. The spectra have been normalized to the flux of the nearest photometric filter. The relative throughput of the FORCAST filters is shown above the data to illustrate the bandwidth.\\}
 \label{fig:sed}
\end{figure*}

In addition to the emission features seen near the systemic velocity, we detect three emission lines that do appear to be associated with multiple peaks. \citet{Lee2007} fit models of the Raman O{\sc vi} $\lambda$6825 line and determined an outflow velocity of the AGB star around 10\,\kms and that the WD hosts a Keplerian disk with an outer rim velocity of 26\,\kms. We detect three emission lines around 5.710--5.725\,$\mu$m that have a velocity-shifted peak at a few \kms and $\sim$30\,\kms. Given the similarity in the predicted and observed velocities, we may be probing the Keplerian disk around the WD. Accretion has been measured in absorption in similar rovibrational water lines in young stellar objects \citep{Indriolo2020} but not before in emission; infrared water absorption bands in the 1--5\,$\mu$m range were also noted by \citet{Bregman1982}. While we do not detect any clear water features in absorption, the disk may not be optically thick because the dust and IR emission appear to be concentrated on the AGB star. We also see one tentative detection of an emission line velocity shifted by $-$76.7\,\kms\ which may be probing the jets of the system.

Several water emission features were detected in both 2021 December, and 2022 April in our 6.7\,$\mu$m setting. These lines are non-gaussian in both epochs and tend to have a systemic peak velocity shift of around 2\,\kms\, except for the line at 6.75107\,$\mu$m, which has a 10\,\kms\ shift between observations. Given our velocity resolution of 0.3\,\kms\ in this setting, we expect the shift is real. We expect that the bulk of the water emission is coming from close-in to the system, $\lesssim$ 70 AU \citep{Gray2022} or 0.\arcsec01 at the distance of HM Sge. These are far smaller scales than those resolvable by EXES (1.\arcsec8--2.\arcsec3). The two settings have position angles that are nearly orthogonal, with the December setting at $\sim$327\degrees (oriented in the northwest direction) and the March section at $\sim$230\degrees (Figure \ref{fig:exes_slit_position}). Given the suspected location of the emission, the difference in positional angle is unlikely to be the cause of the small systemic velocity shift. Further analysis of these data will be conducted in future works.

\subsection{Infrared Observations}
\label{subsec:dust_properties}
The cool component of HM Sge is a thermally pulsing oxygen-rich AGB star, expected to be producing silicate-rich dust. This dust produces a characteristic 10\,$\mu$m silicate feature in the IR that is highly sensitive to changes in dust properties and optical depth. This feature has been seen in HM Sge in several observations, giving us an idea of how the system's dust is evolving.

HM Sge was previously observed in the IR with IRAS, ISO, and VLTI/MIDI. The VLTI observations by \citet{Sacuto2007} showed that mid-IR emission is limited to small spatial scales (half-width at half-maximum $\sim$7.2 mas at 8.5\,$\mu$m). They also found that the local intensity of the envelope at 20 mas is less than 3\% of the intensity at 4 mas, indicating minimal dust near the WD located at 40 mas. Unlike the previous observations in the IR, our FORCAST observations are affected by atmospheric ozone between 9.2 and 10.3\,$\mu$m, shown in gray in Figure \ref{fig:sed}. The new and archival observations show the 10\,$\mu$m silicate feature in emission with varying levels of continuum brightness. The continuum brightness has been steadily decreasing since the initial IRAS observations. Changes in the silicate feature in the FORCAST data are less pronounced but show some subtle differences. In particular, the shape of the full IR continuum has changed, growing brighter in the near-IR and fainter in the far-IR.

These stars can swiftly lose a considerable amount of their circumstellar envelope through dense outflows \citep[up to 10$^{-5}\,M_{\odot}$\,yr$^{-1}$;][]{Hoefner2018}. The mechanism needed to drive away this material is thought to be a dust-driven wind. In this scenario pulsations levitate material out to large radii, where it condenses into dust. This dust and any surrounding gas is then driven away radially by radiation pressure. In less hospitable circumstellar environments it is unclear whether harsher radiation restricts the formation of dust and the AGB mass loss mechanism. The slightly fainter far-IR emission may reflect the relaxing of the system, with the colder dust that resulted from the previous outburst leaving the system. The brighter near-IR emission may also signify that the AGB star is returning to producing dust at full capacity, something we see as additional warm dust.

Within the ISO spectra from 1996 and 1997 we see several emission lines in the near-IR and far-IR that are expected to be emanating from HM Sge's surrounding nebula \citep{Schild2001}. The most prominent is the [Ne {\sc vi}] $^2$P$_{3/2}-^2$P$_{1/2}$ line at 7.6524\,$\mu$m. This line was observed in 1996/97 with ISO and again in 2005/06 with MIDI \citep{Sacuto2007}. In the MIDI observations, the strongest emission was detected in the best-quality spectra, indicating that the line-emitting region is largely extended. The line was not investigated further due to the limited resolution ($R \sim 200$ around $10.6\mu$m). It was also not originally observed by IRAS likely due to the low resolution ($R \sim 14-35$) and larger slit width ($5\arcmin-7.\arcmin5$) compared to ISO ($10-20^{\prime \prime}$). This line is outside the wavelength coverage of our spectroscopic data. The narrow line does, however, fall within our F077 imaging filter, but we do not see a significant increase of the flux in that filter. We also do not see any other clear emission lines in our new FORCAST spectra, indicating that the system is continuing the trend of relaxing after its 1975 outburst.

Previous works have attempted to fit the IR data of HM Sge with radiative transfer models to better understand the system's dust. These have included 1D and 2D spherically symmetric model geometries with single and multiple dust-shell components. These works have found different results, fitting the ISO data with better fits for single-shell \citep{Bogdanov2001} and double-shell 1D dust models \citep{Schild2001,Angeloni2010}. \citet{Sacuto2007} argue that a double-shell dust model would be too extended given the spatial scale of the dust emission in the MIDI observations, and instead support the use of a single-shell model. They also find that this dust distribution shows an increasing level of asymmetry from 8 to 13\,$\mu$m, with a more extended and patchier distribution in the direction perpendicular to our 130$^{\circ}$ binary axis. \\

\subsubsection{1D Radiative Transfer Modeling} \label{subsubsec:Dust_distribution}

\begin{figure}
 \centering
 \includegraphics[width=\linewidth]{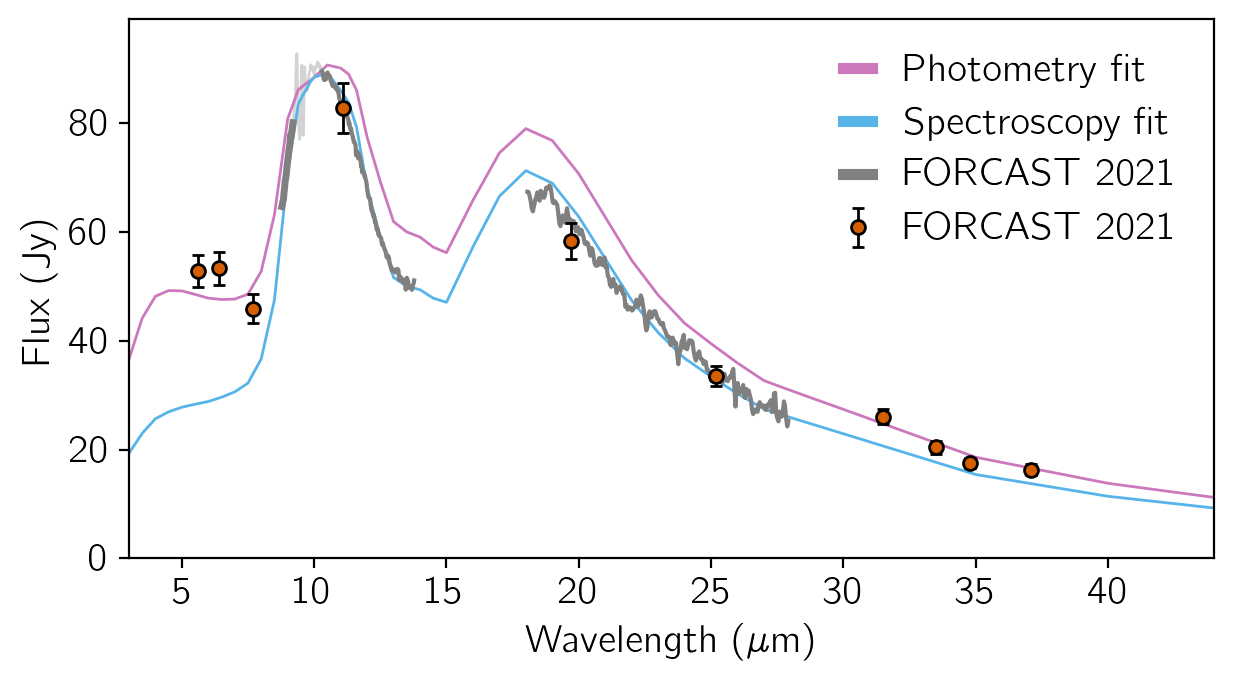}
 \caption{FORCAST photometry and spectra of HM Sge shown with the best-fit radiative transfer models fitted using the \desk; as in Figure \ref{fig:sed}, the region dominated by atmospheric ozone is shown in gray.}
 \label{fig:sed_fits}
\end{figure}

We have fitted the FORCAST data with grids of 1D radiative transfer models from the DUSTY code \citep{Elitzur2001} using the Dusty Evolved Star Kit \citep[{\desk};][]{Goldman2020}. We fit the photometry and spectra using the ``Silicate-Mix'' grids, which include mixtures of different silicate-rich dust species over a range of effective temperatures, inner dust temperatures, and optical depths; the best-fitted models are shown in Figure \ref{fig:sed_fits}. Fitting separately the photometry and spectroscopy results in best-fit models with a luminosity of $L = 1500-2000$\,$L_{\odot}$, lower than the previous estimates based on period-luminosity relations and a distance around two to four times our assumed value \citep[$5,000 - 10,000$\,L$_{\odot}$][]{Bogdanov2001,Schild2001}. We estimate a gas mass loss rate around $\dot M = 4 \times 10^{-6}$\,$M_{\odot}\,{\rm yr}^{-1}$, similar to values previously derived using DUSTY modeling when accounting for the different assumptions for the distance \citep{Bogdanov2001}\footnote{Mass loss rates derived from DUSTY models scale with the luminosity as $\dot{M} \propto r_{gd}^{1/2} L^{3/4}$, where $r_{gd}$ is the gas-to-dust ratio \citep{Elitzur2001}. As the luminosity scales with distance, and assumed distances have differed in previous works, this is expected to affect the comparison of mass loss rates.}. These mass loss rates, however, should be used with caution because incorrect model geometry can have a considerable impact on the derived mass loss rates \citep{Wiegert2020}. Additionally, the system's outburst(s) may result in more-significant episodic mass loss. Regardless, the cool component appears to have swiftly returned to producing significant dust post-outburst. \\

\vspace{0.5cm}

\section{Conclusion} \label{sec:Conclusion}
We have compared new and archival observations in the IR, optical, and UV to understand how the symbiotic Mira HM Sge has evolved on different scales after its 1975 nova-like outburst.

Changes in UV emission lines and narrowband imaging suggest the environment around HM Sge is evolving. The system has shown a significant decrease in the strength of most of the emission lines detected in 1989. This may either be due to gas being ionized to a higher state or be related to a weaker photoionizing flux from an increase in the separation of the two stellar components due to orbital motion. The morphology of the nebular emission closer to the stellar components shows some changes that may be related to rotation or different parts of the nebula changing in brightness.

New and archival observations do not point to any clear changes linked to the AGB outflow or the most recent outburst. We see some movement of nebular features in new and archival UV narrowband images from 1996 to 2021, corresponding to average outflow velocities of 38 and 78\,\kms. We also detect water emission lines in a symbiotic system for the first time, most of which show velocity shifts of a few \kms\ from the systemic velocity. Three new water emission lines are detected with velocity shifts of 31\,\kms, which is consistent with the value (30\,\kms) previously modeled for the outer rim velocity of the accretion disk of the WD.

HM Sge's stellar components have also shown considerable changes over the past few decades. The [Ne\,{\footnotesize V}]$\lambda$1575/[Ne\,{\footnotesize IV}]$\lambda$1601 flux ratio in the UV suggests that the effective temperature of the WD is now at least 250,000~K, compared to less than 200,000~K in 1989. We observe a dimming in the $I$ band over the past year, which may be linked to the orbital motion of the system. Infrared photometry and grism spectroscopy targeting the AGB star show a slightly higher near-IR flux and fainter far-IR flux. This may also be related to the orbital motion or may suggest that the system has returned to producing dust at full capacity. By fitting the IR spectral energy distribution (SED) with radiative transfer models we measure an AGB luminosity of $L = 1500-2000$\,$L_{\odot}$ and a gas mass loss rate of $\dot M = 4 \times 10^{-6}$\,$M_{\odot}\,{\rm yr}^{-1}$. The fairly consistent shape of the AGB SED since the outburst suggests that the AGB star swiftly returned to significant stable dust production.

\begin{acknowledgements}
We would like to thank our anonymous referee for their truly constructive feedback. We would like to thank Curtis Dewitt for his expertise in understanding the EXES data, and Brittany Bennett for illustrating and helping us communicate our work. We would also like to thank Nick Indriolo for his helpful insight, and Romano Corradi for providing the archival NOT data. And last but not least, we would like to thank the observers of the AAVSO for their dedication and passion. Support for program number GO-16492 was provided by NASA through a grant from the Space Telescope Science Institute, which is operated by the Association of Universities for Research in Astronomy, Incorporated, under NASA contract NAS 526555. Support for this work was also provided by NASA through award N2216 issued by USRA. This research is based on observations made with the NASA/ESA Hubble Space Telescope obtained from the Space Telescope Science Institute, which is operated by the Association of Universities for Research in Astronomy, Inc., under NASA contract NAS 5–26555. These observations are associated with program GO-16492. This work is also based in part on observations made with the NASA/DLR Stratospheric Observatory for Infrared Astronomy (SOFIA). SOFIA is jointly operated by the Universities Space Research Association, Inc. (USRA), under NASA contract NNA17BF53C, and the Deutsches SOFIA Institut (DSI) under DLR contract 50 OK 0901 to the University of Stuttgart.
\end{acknowledgements}

\facilities{\hst, SOFIA}
\software{DESK \citep{Goldman2020}, Astropy \citep{astropy:2013, astropy:2018}, Matplotlib \citep{matplotlib2007}, Scipy \citep{scipy2020}, Numpy \citep{numpy2011,numpy2020}, IPython \citep{ipython2007}, pandas \citep{pandas}, Astroalign \citep{Astroalign}, Photutils \citep{photutils}.}

\bibliography{references_2020}{}

\begin{thebibliography}{}
\expandafter\ifx\csname natexlab\endcsname\relax\def\natexlab#1{#1}\fi
\providecommand{\url}[1]{\href{#1}{#1}}
\providecommand{\dodoi}[1]{doi:~\href{http://doi.org/#1}{\nolinkurl{#1}}}
\providecommand{\doeprint}[1]{\href{http://ascl.net/#1}{\nolinkurl{http://ascl.net/#1}}}
\providecommand{\doarXiv}[1]{\href{https://arxiv.org/abs/#1}{\nolinkurl{https://arxiv.org/abs/#1}}}

\bibitem[{{Andriantsaralaza} {et~al.}(2022){Andriantsaralaza}, {Ramstedt},
  {Vlemmings}, \& {De Beck}}]{Andriantsaralaza2022}
{Andriantsaralaza}, M., {Ramstedt}, S., {Vlemmings}, W.~H.~T., \& {De Beck}, E.
  2022, \aap, 667, A74, \dodoi{10.1051/0004-6361/202243670}

\bibitem[{{Angeloni} {et~al.}(2010){Angeloni}, {Contini}, {Ciroi}, \&
  {Rafanelli}}]{Angeloni2010}
{Angeloni}, R., {Contini}, M., {Ciroi}, S., \& {Rafanelli}, P. 2010, \mnras,
  402, 2075, \dodoi{10.1111/j.1365-2966.2009.16067.x}

\bibitem[{{Astropy Collaboration} {et~al.}(2013){Astropy Collaboration},
  {Robitaille}, {Tollerud}, {Greenfield}, {Droettboom}, {Bray}, {Aldcroft},
  {Davis}, {Ginsburg}, {Price-Whelan}, {Kerzendorf}, {Conley}, {Crighton},
  {Barbary}, {Muna}, {Ferguson}, {Grollier}, {Parikh}, {Nair}, {Unther},
  {Deil}, {Woillez}, {Conseil}, {Kramer}, {Turner}, {Singer}, {Fox}, {Weaver},
  {Zabalza}, {Edwards}, {Azalee Bostroem}, {Burke}, {Casey}, {Crawford},
  {Dencheva}, {Ely}, {Jenness}, {Labrie}, {Lim}, {Pierfederici}, {Pontzen},
  {Ptak}, {Refsdal}, {Servillat}, \& {Streicher}}]{astropy:2013}
{Astropy Collaboration}, {Robitaille}, T.~P., {Tollerud}, E.~J., {et~al.} 2013,
  \aap, 558, A33, \dodoi{10.1051/0004-6361/201322068}

\bibitem[{{Astropy Collaboration} {et~al.}(2018){Astropy Collaboration},
  {Price-Whelan}, {Sip{\H{o}}cz}, {G{\"u}nther}, {Lim}, {Crawford}, {Conseil},
  {Shupe}, {Craig}, {Dencheva}, {Ginsburg}, {Vand erPlas}, {Bradley},
  {P{\'e}rez-Su{\'a}rez}, {de Val-Borro}, {Aldcroft}, {Cruz}, {Robitaille},
  {Tollerud}, {Ardelean}, {Babej}, {Bach}, {Bachetti}, {Bakanov}, {Bamford},
  {Barentsen}, {Barmby}, {Baumbach}, {Berry}, {Biscani}, {Boquien}, {Bostroem},
  {Bouma}, {Brammer}, {Bray}, {Breytenbach}, {Buddelmeijer}, {Burke},
  {Calderone}, {Cano Rodr{\'\i}guez}, {Cara}, {Cardoso}, {Cheedella}, {Copin},
  {Corrales}, {Crichton}, {D'Avella}, {Deil}, {Depagne}, {Dietrich}, {Donath},
  {Droettboom}, {Earl}, {Erben}, {Fabbro}, {Ferreira}, {Finethy}, {Fox},
  {Garrison}, {Gibbons}, {Goldstein}, {Gommers}, {Greco}, {Greenfield},
  {Groener}, {Grollier}, {Hagen}, {Hirst}, {Homeier}, {Horton}, {Hosseinzadeh},
  {Hu}, {Hunkeler}, {Ivezi{\'c}}, {Jain}, {Jenness}, {Kanarek}, {Kendrew},
  {Kern}, {Kerzendorf}, {Khvalko}, {King}, {Kirkby}, {Kulkarni}, {Kumar},
  {Lee}, {Lenz}, {Littlefair}, {Ma}, {Macleod}, {Mastropietro}, {McCully},
  {Montagnac}, {Morris}, {Mueller}, {Mumford}, {Muna}, {Murphy}, {Nelson},
  {Nguyen}, {Ninan}, {N{\"o}the}, {Ogaz}, {Oh}, {Parejko}, {Parley}, {Pascual},
  {Patil}, {Patil}, {Plunkett}, {Prochaska}, {Rastogi}, {Reddy Janga},
  {Sabater}, {Sakurikar}, {Seifert}, {Sherbert}, {Sherwood-Taylor}, {Shih},
  {Sick}, {Silbiger}, {Singanamalla}, {Singer}, {Sladen}, {Sooley},
  {Sornarajah}, {Streicher}, {Teuben}, {Thomas}, {Tremblay}, {Turner},
  {Terr{\'o}n}, {van Kerkwijk}, {de la Vega}, {Watkins}, {Weaver}, {Whitmore},
  {Woillez}, {Zabalza}, \& {Astropy Contributors}}]{astropy:2018}
{Astropy Collaboration}, {Price-Whelan}, A.~M., {Sip{\H{o}}cz}, B.~M., {et~al.}
  2018, \aj, 156, 123, \dodoi{10.3847/1538-3881/aabc4f}

\bibitem[{{Belczy{\'n}ski} {et~al.}(2000){Belczy{\'n}ski}, {Miko{\l}ajewska},
  {Munari}, {Ivison}, \& {Friedjung}}]{Belczynski2000}
{Belczy{\'n}ski}, K., {Miko{\l}ajewska}, J., {Munari}, U., {Ivison}, R.~J., \&
  {Friedjung}, M. 2000, \aaps, 146, 407, \dodoi{10.1051/aas:2000280}

\bibitem[{{Beliakina} {et~al.}(1978){Beliakina}, {Gershberg}, \&
  {Shakovskaia}}]{Beliakina1978}
{Beliakina}, T.~S., {Gershberg}, R.~E., \& {Shakovskaia}, N.~I. 1978, Soviet
  Astronomy Letters, 4, 219

\bibitem[{{Belyakina} {et~al.}(1979){Belyakina}, {Gershberg}, \&
  {Schakhovskaya}}]{Belyakina1979}
{Belyakina}, T.~S., {Gershberg}, R.~E., \& {Schakhovskaya}, N.~I. 1979, Soviet
  Astronomy Letters, 5, 349

\bibitem[{Beroiz {et~al.}(2020)Beroiz, Cabral, \& Sanchez}]{Astroalign}
Beroiz, M., Cabral, J., \& Sanchez, B. 2020, Astronomy and Computing, 32,
  100384, \dodoi{https://doi.org/10.1016/j.ascom.2020.100384}

\bibitem[{{Bogdanov} \& {Taranova}(2001)}]{Bogdanov2001}
{Bogdanov}, M.~B., \& {Taranova}, O.~G. 2001, Astronomy Reports, 45, 44,
  \dodoi{10.1134/1.1336600}

\bibitem[{Bradley {et~al.}(2020)Bradley, Sip{\H o}cz, Robitaille, Tollerud,
  Vin{\'{\i}}cius, Deil, Barbary, Wilson, Busko, G{\"u}nther, Cara, Conseil,
  Bostroem, Droettboom, Bray, Bratholm, Lim, Barentsen, Craig, Pascual, Perren,
  Greco, Donath, de~Val-Borro, Kerzendorf, Bach, Weaver, D'Eugenio, Souchereau,
  \& Ferreira}]{photutils}
Bradley, L., Sip{\H o}cz, B., Robitaille, T., {et~al.} 2020, astropy/photutils:
  1.0.0, 1.0.0,  Zenodo, \dodoi{10.5281/zenodo.4044744}

\bibitem[{{Bregman}(1982)}]{Bregman1982}
{Bregman}, J.~D. 1982, in Bulletin of the American Astronomical Society,
  Vol.~14, 982

\bibitem[{{Cho} \& {Kim}(2010)}]{Cho2010}
{Cho}, S.-H., \& {Kim}, J. 2010, \apj, 719, 126,
  \dodoi{10.1088/0004-637X/719/1/126}

\bibitem[{{Ciatti} {et~al.}(1979){Ciatti}, {Mammano}, \&
  {Vittone}}]{Ciatti1979}
{Ciatti}, F., {Mammano}, A., \& {Vittone}, A. 1979, \aap, 79, 247

\bibitem[{{Corradi} {et~al.}(1999){Corradi}, {Ferrer}, {Schwarz}, {Brand i}, \&
  {Garc{\'\i}a}}]{Corradi1999}
{Corradi}, R.~L.~M., {Ferrer}, O.~E., {Schwarz}, H.~E., {Brand i}, E., \&
  {Garc{\'\i}a}, L. 1999, \aap, 348, 978

\bibitem[{{Dokuchaeva}(1976)}]{Dokuchaeva1976}
{Dokuchaeva}, O.~D. 1976, Information Bulletin on Variable Stars, 1189, 1

\bibitem[{{Elitzur} \& {Ivezi{\'c}}(2001)}]{Elitzur2001}
{Elitzur}, M., \& {Ivezi{\'c}}, {\v{Z}}. 2001, \mnras, 327, 403,
  \dodoi{10.1046/j.1365-8711.2001.04706.x}

\bibitem[{{Engels} \& {Bunzel}(2015)}]{Engels2015}
{Engels}, D., \& {Bunzel}, F. 2015, \aap, 582, A68,
  \dodoi{10.1051/0004-6361/201322589}

\bibitem[{{Eyres} {et~al.}(2001){Eyres}, {Bode}, {Taylor}, {Crocker}, \&
  {Davis}}]{Eyres2001}
{Eyres}, S.~P.~S., {Bode}, M.~F., {Taylor}, A.~R., {Crocker}, M.~M., \&
  {Davis}, R.~J. 2001, \apj, 551, 512, \dodoi{10.1086/320086}

\bibitem[{{Fonfr{\'\i}a} {et~al.}(2008){Fonfr{\'\i}a}, {Cernicharo}, {Richter},
  \& {Lacy}}]{Fonfria2008}
{Fonfr{\'\i}a}, J.~P., {Cernicharo}, J., {Richter}, M.~J., \& {Lacy}, J.~H.
  2008, \apj, 673, 445, \dodoi{10.1086/523882}

\bibitem[{{Formiggini} {et~al.}(1995){Formiggini}, {Contini}, \&
  {Leibowitz}}]{Formiggini1995}
{Formiggini}, L., {Contini}, M., \& {Leibowitz}, E.~M. 1995, \mnras, 277, 1071,
  \dodoi{10.1093/mnras/277.3.1071}

\bibitem[{{Gaia Collaboration} {et~al.}(2018){Gaia Collaboration}, {Brown},
  {Vallenari}, {Prusti}, {de Bruijne}, {Babusiaux}, {Bailer-Jones}, {Biermann},
  {Evans}, {Eyer}, {Jansen}, {Jordi}, {Klioner}, {Lammers}, {Lindegren},
  {Luri}, {Mignard}, {Panem}, {Pourbaix}, {Randich}, {Sartoretti}, {Siddiqui},
  {Soubiran}, {van Leeuwen}, {Walton}, {Arenou}, {Bastian}, {Cropper},
  {Drimmel}, {Katz}, {Lattanzi}, {Bakker}, {Cacciari}, {Casta{\~n}eda},
  {Chaoul}, {Cheek}, {De Angeli}, {Fabricius}, {Guerra}, {Holl}, {Masana},
  {Messineo}, {Mowlavi}, {Nienartowicz}, {Panuzzo}, {Portell}, {Riello},
  {Seabroke}, {Tanga}, {Th{\'e}venin}, {Gracia-Abril}, {Comoretto},
  {Garcia-Reinaldos}, {Teyssier}, {Altmann}, {Andrae}, {Audard},
  {Bellas-Velidis}, {Benson}, {Berthier}, {Blomme}, {Burgess}, {Busso},
  {Carry}, {Cellino}, {Clementini}, {Clotet}, {Creevey}, {Davidson}, {De
  Ridder}, {Delchambre}, {Dell'Oro}, {Ducourant},
  {Fern{\'a}ndez-Hern{\'a}ndez}, {Fouesneau}, {Fr{\'e}mat}, {Galluccio},
  {Garc{\'\i}a-Torres}, {Gonz{\'a}lez-N{\'u}{\~n}ez}, {Gonz{\'a}lez-Vidal},
  {Gosset}, {Guy}, {Halbwachs}, {Hambly}, {Harrison}, {Hern{\'a}ndez},
  {Hestroffer}, {Hodgkin}, {Hutton}, {Jasniewicz}, {Jean-Antoine-Piccolo},
  {Jordan}, {Korn}, {Krone-Martins}, {Lanzafame}, {Lebzelter}, {L{\"o}ffler},
  {Manteiga}, {Marrese}, {Mart{\'\i}n-Fleitas}, {Moitinho}, {Mora}, {Muinonen},
  {Osinde}, {Pancino}, {Pauwels}, {Petit}, {Recio-Blanco}, {Richards},
  {Rimoldini}, {Robin}, {Sarro}, {Siopis}, {Smith}, {Sozzetti}, {S{\"u}veges},
  {Torra}, {van Reeven}, {Abbas}, {Abreu Aramburu}, {Accart}, {Aerts},
  {Altavilla}, {{\'A}lvarez}, {Alvarez}, {Alves}, {Anderson}, {Andrei},
  {Anglada Varela}, {Antiche}, {Antoja}, {Arcay}, {Astraatmadja}, {Bach},
  {Baker}, {Balaguer-N{\'u}{\~n}ez}, {Balm}, {Barache}, {Barata}, {Barbato},
  {Barblan}, {Barklem}, {Barrado}, {Barros}, {Barstow}, {Bartholom{\'e}
  Mu{\~n}oz}, {Bassilana}, {Becciani}, {Bellazzini}, {Berihuete}, {Bertone},
  {Bianchi}, {Bienaym{\'e}}, {Blanco-Cuaresma}, {Boch}, {Boeche}, {Bombrun},
  {Borrachero}, {Bossini}, {Bouquillon}, {Bourda}, {Bragaglia}, {Bramante},
  {Breddels}, {Bressan}, {Brouillet}, {Br{\"u}semeister}, {Brugaletta},
  {Bucciarelli}, {Burlacu}, {Busonero}, {Butkevich}, {Buzzi}, {Caffau},
  {Cancelliere}, {Cannizzaro}, {Cantat-Gaudin}, {Carballo}, {Carlucci},
  {Carrasco}, {Casamiquela}, {Castellani}, {Castro-Ginard}, {Charlot},
  {Chemin}, {Chiavassa}, {Cocozza}, {Costigan}, {Cowell}, {Crifo}, {Crosta},
  {Crowley}, {Cuypers}, {Dafonte}, {Damerdji}, {Dapergolas}, {David}, {David},
  {de Laverny}, {De Luise}, {De March}, {de Martino}, {de Souza}, {de Torres},
  {Debosscher}, {del Pozo}, {Delbo}, {Delgado}, {Delgado}, {Di Matteo},
  {Diakite}, {Diener}, {Distefano}, {Dolding}, {Drazinos}, {Dur{\'a}n},
  {Edvardsson}, {Enke}, {Eriksson}, {Esquej}, {Eynard Bontemps}, {Fabre},
  {Fabrizio}, {Faigler}, {Falc{\~a}o}, {Farr{\`a}s Casas}, {Federici},
  {Fedorets}, {Fernique}, {Figueras}, {Filippi}, {Findeisen}, {Fonti},
  {Fraile}, {Fraser}, {Fr{\'e}zouls}, {Gai}, {Galleti}, {Garabato},
  {Garc{\'\i}a-Sedano}, {Garofalo}, {Garralda}, {Gavel}, {Gavras}, {Gerssen},
  {Geyer}, {Giacobbe}, {Gilmore}, {Girona}, {Giuffrida}, {Glass}, {Gomes},
  {Granvik}, {Gueguen}, {Guerrier}, {Guiraud}, {Guti{\'e}rrez-S{\'a}nchez},
  {Haigron}, {Hatzidimitriou}, {Hauser}, {Haywood}, {Heiter}, {Helmi}, {Heu},
  {Hilger}, {Hobbs}, {Hofmann}, {Holland}, {Huckle}, {Hypki}, {Icardi},
  {Jan{\ss}en}, {Jevardat de Fombelle}, {Jonker}, {Juh{\'a}sz}, {Julbe},
  {Karampelas}, {Kewley}, {Klar}, {Kochoska}, {Kohley}, {Kolenberg},
  {Kontizas}, {Kontizas}, {Koposov}, {Kordopatis}, {Kostrzewa-Rutkowska},
  {Koubsky}, {Lambert}, {Lanza}, {Lasne}, {Lavigne}, {Le Fustec}, {Le
  Poncin-Lafitte}, {Lebreton}, {Leccia}, {Leclerc}, {Lecoeur-Taibi},
  {Lenhardt}, {Leroux}, {Liao}, {Licata}, {Lindstr{\o}m}, {Lister}, {Livanou},
  {Lobel}, {L{\'o}pez}, {Managau}, {Mann}, {Mantelet}, {Marchal}, {Marchant},
  {Marconi}, {Marinoni}, {Marschalk{\'o}}, {Marshall}, {Martino}, {Marton},
  {Mary}, {Massari}, {Matijevi{\v{c}}}, {Mazeh}, {McMillan}, {Messina},
  {Michalik}, {Millar}, {Molina}, {Molinaro}, {Moln{\'a}r}, {Montegriffo},
  {Mor}, {Morbidelli}, {Morel}, {Morris}, {Mulone}, {Muraveva}, {Musella},
  {Nelemans}, {Nicastro}, {Noval}, {O'Mullane}, {Ord{\'e}novic},
  {Ord{\'o}{\~n}ez-Blanco}, {Osborne}, {Pagani}, {Pagano}, {Pailler},
  {Palacin}, {Palaversa}, {Panahi}, {Pawlak}, {Piersimoni}, {Pineau}, {Plachy},
  {Plum}, {Poggio}, {Poujoulet}, {Pr{\v{s}}a}, {Pulone}, {Racero}, {Ragaini},
  {Rambaux}, {Ramos-Lerate}, {Regibo}, {Reyl{\'e}}, {Riclet}, {Ripepi}, {Riva},
  {Rivard}, {Rixon}, {Roegiers}, {Roelens}, {Romero-G{\'o}mez}, {Rowell},
  {Royer}, {Ruiz-Dern}, {Sadowski}, {Sagrist{\`a} Sell{\'e}s}, {Sahlmann},
  {Salgado}, {Salguero}, {Sanna}, {Santana-Ros}, {Sarasso}, {Savietto},
  {Schultheis}, {Sciacca}, {Segol}, {Segovia}, {S{\'e}gransan}, {Shih},
  {Siltala}, {Silva}, {Smart}, {Smith}, {Solano}, {Solitro}, {Sordo}, {Soria
  Nieto}, {Souchay}, {Spagna}, {Spoto}, {Stampa}, {Steele},
  {Steidelm{\"u}ller}, {Stephenson}, {Stoev}, {Suess}, {Surdej}, {Szabados},
  {Szegedi-Elek}, {Tapiador}, {Taris}, {Tauran}, {Taylor}, {Teixeira},
  {Terrett}, {Teyssandier}, {Thuillot}, {Titarenko}, {Torra Clotet}, {Turon},
  {Ulla}, {Utrilla}, {Uzzi}, {Vaillant}, {Valentini}, {Valette}, {van Elteren},
  {Van Hemelryck}, {van Leeuwen}, {Vaschetto}, {Vecchiato}, {Veljanoski},
  {Viala}, {Vicente}, {Vogt}, {von Essen}, {Voss}, {Votruba}, {Voutsinas},
  {Walmsley}, {Weiler}, {Wertz}, {Wevers}, {Wyrzykowski}, {Yoldas},
  {{\v{Z}}erjal}, {Ziaeepour}, {Zorec}, {Zschocke}, {Zucker}, {Zurbach}, \&
  {Zwitter}}]{Gaia2018}
{Gaia Collaboration}, {Brown}, A.~G.~A., {Vallenari}, A., {et~al.} 2018, \aap,
  616, A1, \dodoi{10.1051/0004-6361/201833051}

\bibitem[{{Girard} \& {Willson}(1987)}]{Girard1987}
{Girard}, T., \& {Willson}, L.~A. 1987, \aap, 183, 247

\bibitem[{Goldman(2020)}]{Goldman2020}
Goldman, S.~R. 2020, Journal of Open Source Software, 5, 2554,
  \dodoi{10.21105/joss.02554}

\bibitem[{Goldman {et~al.}(2022)Goldman, Sankrit, Wolthuis, Garner, Gualdoni,
  \& Bolzoni}]{Goldman_2022_research_note}
Goldman, S.~R., Sankrit, R., Wolthuis, N., {et~al.} 2022, Research Notes of the
  {AAS}, 6, 159, \dodoi{10.3847/2515-5172/ac8808}

\bibitem[{{Goldman} {et~al.}(2017){Goldman}, {van Loon}, {Zijlstra}, {Green},
  {Wood}, {Nanni}, {Imai}, {Whitelock}, {Matsuura}, {Groenewegen}, \&
  {G{\'o}mez}}]{Goldman2017}
{Goldman}, S.~R., {van Loon}, J.~T., {Zijlstra}, A.~A., {et~al.} 2017, \mnras,
  465, 403, \dodoi{10.1093/mnras/stw2708}

\bibitem[{{Gray} {et~al.}(2022){Gray}, {Etoka}, {Richards}, \&
  {Pimpanuwat}}]{Gray2022}
{Gray}, M.~D., {Etoka}, S., {Richards}, A.~M.~S., \& {Pimpanuwat}, B. 2022,
  \mnras, 513, 1354, \dodoi{10.1093/mnras/stac854}

\bibitem[{{Guerrero} {et~al.}(2020){Guerrero}, {Suzett Rechy-Garc{\'\i}a}, \&
  {Ortiz}}]{Guerrero2020}
{Guerrero}, M.~A., {Suzett Rechy-Garc{\'\i}a}, J., \& {Ortiz}, R. 2020, \apj,
  890, 50, \dodoi{10.3847/1538-4357/ab61fa}

\bibitem[{{Hack} \& {Paresce}(1993)}]{Hack1993}
{Hack}, W.~J., \& {Paresce}, F. 1993, \pasp, 105, 1273, \dodoi{10.1086/133307}

\bibitem[{{Harris} {et~al.}(2020){Harris}, {Millman}, {van der Walt},
  {Gommers}, {Virtanen}, {Cournapeau}, {Wieser}, {Taylor}, {Berg}, {Smith},
  {Kern}, {Picus}, {Hoyer}, {van Kerkwijk}, {Brett}, {Haldane}, {del R{\'\i}o},
  {Wiebe}, {Peterson}, {G{\'e}rard-Marchant}, {Sheppard}, {Reddy}, {Weckesser},
  {Abbasi}, {Gohlke}, \& {Oliphant}}]{numpy2020}
{Harris}, C.~R., {Millman}, K.~J., {van der Walt}, S.~J., {et~al.} 2020, \nat,
  585, 357, \dodoi{10.1038/s41586-020-2649-2}

\bibitem[{{Herter} {et~al.}(2018){Herter}, {Adams}, {Gull}, {Schoenwald},
  {Keller}, {Pirger}, {Henderson}, {Stacey}, {Nikola}, {De Buizer}, {Vacca}, \&
  {Ennico}}]{Herter2018}
{Herter}, T.~L., {Adams}, J.~D., {Gull}, G.~E., {et~al.} 2018, Journal of
  Astronomical Instrumentation, 7, 1840005, \dodoi{10.1142/S2251171718400056}

\bibitem[{{Hinkle} {et~al.}(2013){Hinkle}, {Fekel}, {Joyce}, \&
  {Wood}}]{Hinkle2013}
{Hinkle}, K.~H., {Fekel}, F.~C., {Joyce}, R.~R., \& {Wood}, P. 2013, \apj, 770,
  28, \dodoi{10.1088/0004-637X/770/1/28}

\bibitem[{{H{\"o}fner} \& {Olofsson}(2018)}]{Hoefner2018}
{H{\"o}fner}, S., \& {Olofsson}, H. 2018, \aapr, 26, 1,
  \dodoi{10.1007/s00159-017-0106-5}

\bibitem[{Hunter(2007)}]{matplotlib2007}
Hunter, J.~D. 2007, Computing in Science Engineering, 9, 90,
  \dodoi{10.1109/MCSE.2007.55}

\bibitem[{{Indriolo} {et~al.}(2020){Indriolo}, {Neufeld}, {Barr}, {Boogert},
  {DeWitt}, {Karska}, {Montiel}, {Richter}, \& {Tielens}}]{Indriolo2020}
{Indriolo}, N., {Neufeld}, D.~A., {Barr}, A.~G., {et~al.} 2020, \apj, 894, 107,
  \dodoi{10.3847/1538-4357/ab88a1}

\bibitem[{Kenyon(1986)}]{Kenyon1986}
Kenyon, S. 1986, Cambridge and New York, Cambridge University Press, 1986, 295
  p., -1, \dodoi{10.1017/CBO9780511586071}

\bibitem[{{Kwok} {et~al.}(1984){Kwok}, {Bignell}, \& {Purton}}]{Kwok1984}
{Kwok}, S., {Bignell}, R.~C., \& {Purton}, C.~R. 1984, \apj, 279, 188,
  \dodoi{10.1086/161881}

\bibitem[{{Kwok} \& {Leahy}(1984)}]{Kwok_Leahy_1984}
{Kwok}, S., \& {Leahy}, D.~A. 1984, \apj, 283, 675, \dodoi{10.1086/162353}

\bibitem[{{Lee} \& {Kang}(2007)}]{Lee2007}
{Lee}, H.-W., \& {Kang}, S. 2007, \apj, 669, 1156, \dodoi{10.1086/521719}

\bibitem[{{Liimets} {et~al.}(2018){Liimets}, {Corradi}, {Jones}, {Verro},
  {Santander-Garc{\'\i}a}, {Kolka}, {Sidonio}, {Kankare}, {Kankare}, {Pursimo},
  \& {Wilson}}]{Liimets2018}
{Liimets}, T., {Corradi}, R.~L.~M., {Jones}, D., {et~al.} 2018, \aap, 612,
  A118, \dodoi{10.1051/0004-6361/201732073}

\bibitem[{{Merc} {et~al.}(2019){Merc}, {G{\'a}lis}, \& {Wolf}}]{Merc2019}
{Merc}, J., {G{\'a}lis}, R., \& {Wolf}, M. 2019, Research Notes of the American
  Astronomical Society, 3, 28, \dodoi{10.3847/2515-5172/ab0429}

\bibitem[{Merrill(1940)}]{Merrill1940}
Merrill, P.~W. 1940, Spectra of long-period variable stars (Chicago, IL: Univ.
  Chicago Press).
\newblock \url{http://adsabs.harvard.edu/abs/1940slpv.book.....M}

\bibitem[{{Mueller} \& {Nussbaumer}(1985)}]{Mueller1985}
{Mueller}, B.~E.~A., \& {Nussbaumer}, H. 1985, \aap, 145, 144

\bibitem[{{Muerset} {et~al.}(1997){Muerset}, {Wolff}, \&
  {Jordan}}]{Muerset1997}
{Muerset}, U., {Wolff}, B., \& {Jordan}, S. 1997, \aap, 319, 201

\bibitem[{{Munari} \& {Whitelock}(1989)}]{Munari1989}
{Munari}, U., \& {Whitelock}, P.~A. 1989, \mnras, 237, 45P,
  \dodoi{10.1093/mnras/237.1.45P}

\bibitem[{{Murset} \& {Nussbaumer}(1994)}]{Murset1994}
{Murset}, U., \& {Nussbaumer}, H. 1994, \aap, 282, 586

\bibitem[{{M{\"u}rset} \& {Schmid}(1999)}]{Murset1999}
{M{\"u}rset}, U., \& {Schmid}, H.~M. 1999, \aaps, 137, 473,
  \dodoi{10.1051/aas:1999105}

\bibitem[{{Neugebauer} {et~al.}(1984){Neugebauer}, {Habing}, {van Duinen},
  {Aumann}, {Baud}, {Beichman}, {Beintema}, {Boggess}, {Clegg}, {de Jong},
  {Emerson}, {Gautier}, {Gillett}, {Harris}, {Hauser}, {Houck}, {Jennings},
  {Low}, {Marsden}, {Miley}, {Olnon}, {Pottasch}, {Raimond}, {Rowan-Robinson},
  {Soifer}, {Walker}, {Wesselius}, \& {Young}}]{Neugebauer1984}
{Neugebauer}, G., {Habing}, H.~J., {van Duinen}, R., {et~al.} 1984, \apjl, 278,
  L1, \dodoi{10.1086/184209}

\bibitem[{{Nussbaumer} \& {Vogel}(1990)}]{Nussbaumer1990}
{Nussbaumer}, H., \& {Vogel}, M. 1990, \aap, 236, 117

\bibitem[{{Olofsson} {et~al.}(2022){Olofsson}, {Khouri}, {Sargent}, {Winnberg},
  {Blommaert}, {Groenewegen}, {Muller}, {Kastner}, {Meixner}, {Otsuka},
  {Patel}, {Ryde}, \& {Srinivasan}}]{Olofsson2022}
{Olofsson}, H., {Khouri}, T., {Sargent}, B.~A., {et~al.} 2022, \aap, 665, A82,
  \dodoi{10.1051/0004-6361/202244053}

\bibitem[{{Palen} {et~al.}(2002){Palen}, {Balick}, {Hajian}, {Terzian}, {Bond},
  \& {Panagia}}]{Palen2002}
{Palen}, S., {Balick}, B., {Hajian}, A.~R., {et~al.} 2002, \aj, 123, 2666,
  \dodoi{10.1086/339838}

\bibitem[{pandas~development team(2022)}]{pandas}
pandas~development team, T. 2022, pandas-dev/pandas: Pandas 1.4.3, v1.4.3,
  Zenodo, \dodoi{10.5281/zenodo.6702671}

\bibitem[{Perez \& Granger(2007)}]{ipython2007}
Perez, F., \& Granger, B.~E. 2007, Computing in Science Engineering, 9, 21,
  \dodoi{10.1109/MCSE.2007.53}

\bibitem[{{Reed} {et~al.}(1999){Reed}, {Balick}, {Hajian}, {Klayton},
  {Giovanardi}, {Casertano}, {Panagia}, \& {Terzian}}]{Reed1999}
{Reed}, D.~S., {Balick}, B., {Hajian}, A.~R., {et~al.} 1999, \aj, 118, 2430,
  \dodoi{10.1086/301091}

\bibitem[{{Richards} {et~al.}(1999){Richards}, {Bode}, {Eyres}, {Kenny},
  {Davis}, \& {Watson}}]{Richards1999}
{Richards}, A.~M.~S., {Bode}, M.~F., {Eyres}, S.~P.~S., {et~al.} 1999, \mnras,
  305, 380, \dodoi{10.1046/j.1365-8711.1999.02465.x}

\bibitem[{{Richter} {et~al.}(2018){Richter}, {Dewitt}, {McKelvey}, {Montiel},
  {McMurray}, \& {Case}}]{Richter2018}
{Richter}, M.~J., {Dewitt}, C.~N., {McKelvey}, M., {et~al.} 2018, Journal of
  Astronomical Instrumentation, 7, 1840013, \dodoi{10.1142/S2251171718400135}

\bibitem[{{Sacuto} {et~al.}(2007){Sacuto}, {Chesneau}, {Vannier}, \&
  {Cruzal{\`e}bes}}]{Sacuto2007}
{Sacuto}, S., {Chesneau}, O., {Vannier}, M., \& {Cruzal{\`e}bes}, P. 2007,
  \aap, 465, 469, \dodoi{10.1051/0004-6361:20066642}

\bibitem[{{Sahai} {et~al.}(2022){Sahai}, {Huang}, {Scibelli}, {Morris},
  {Hinkle}, \& {Lee}}]{Sahai2022}
{Sahai}, R., {Huang}, P.~S., {Scibelli}, S., {et~al.} 2022, \apj, 929, 59,
  \dodoi{10.3847/1538-4357/ac568a}

\bibitem[{{Schild} {et~al.}(2001){Schild}, {Eyres}, {Salama}, \&
  {Evans}}]{Schild2001}
{Schild}, H., {Eyres}, S.~P.~S., {Salama}, A., \& {Evans}, A. 2001, \aap, 378,
  146, \dodoi{10.1051/0004-6361:20011155}

\bibitem[{{Schmid} {et~al.}(2000){Schmid}, {Corradi}, {Krautter}, \&
  {Schild}}]{Schmid2000}
{Schmid}, H.~M., {Corradi}, R., {Krautter}, J., \& {Schild}, H. 2000, \aap,
  355, 261

\bibitem[{{Solf}(1983)}]{Solf1983}
{Solf}, J. 1983, \apjl, 266, L113, \dodoi{10.1086/183989}

\bibitem[{{Solf}(1984)}]{Solf1984}
---. 1984, \aap, 139, 296

\bibitem[{{Stauffer}(1984)}]{Stauffer1984}
{Stauffer}, J.~R. 1984, \apj, 280, 695, \dodoi{10.1086/162043}

\bibitem[{{Taranova} \& {Shenavrin}(2000)}]{Taranova2000}
{Taranova}, O.~G., \& {Shenavrin}, V.~I. 2000, Astronomy Letters, 26, 600,
  \dodoi{10.1134/1.1307894}

\bibitem[{{Temi} {et~al.}(2018){Temi}, {Hoffman}, {Ennico}, \& {Le}}]{Temi2018}
{Temi}, P., {Hoffman}, D., {Ennico}, K., \& {Le}, J. 2018, Journal of
  Astronomical Instrumentation, 7, 1840011, \dodoi{10.1142/S2251171718400111}

\bibitem[{{Toal{\'a}} {et~al.}(2023){Toal{\'a}}, {Botello}, \&
  {Sabin}}]{Toala2023}
{Toal{\'a}}, J.~A., {Botello}, M.~K., \& {Sabin}, L. 2023, \apj, 948, 14,
  \dodoi{10.3847/1538-4357/acc659}

\bibitem[{{Tomov} {et~al.}(2007){Tomov}, {Tomova}, \& {Bisikalo}}]{Tomov2007}
{Tomov}, N.~A., {Tomova}, M.~T., \& {Bisikalo}, D.~V. 2007, \mnras, 376, L16,
  \dodoi{10.1111/j.1745-3933.2007.00277.x}

\bibitem[{{Tomov} {et~al.}(2017){Tomov}, {Zamanov}, {Ga{\l}an}, \&
  {Pietrukowicz}}]{Tomov2017}
{Tomov}, T., {Zamanov}, R., {Ga{\l}an}, C., \& {Pietrukowicz}, P. 2017, \actaa,
  67, 225, \dodoi{10.32023/0001-5237/67.3.2}

\bibitem[{{van der Walt} {et~al.}(2011){van der Walt}, {Colbert}, \&
  {Varoquaux}}]{numpy2011}
{van der Walt}, S., {Colbert}, S.~C., \& {Varoquaux}, G. 2011, Computing in
  Science and Engineering, 13, 22, \dodoi{10.1109/MCSE.2011.37}

\bibitem[{Virtanen {et~al.}(2020)Virtanen, Gommers, Oliphant, Haberland, Reddy,
  Cournapeau, Burovski, Peterson, Weckesser, Bright, {van der Walt}, Brett,
  Wilson, Millman, Mayorov, Nelson, Jones, Kern, Larson, Carey, Polat, Feng,
  Moore, {VanderPlas}, Laxalde, Perktold, Cimrman, Henriksen, Quintero, Harris,
  Archibald, Ribeiro, Pedregosa, {van Mulbregt}, \& {SciPy 1.0
  Contributors}}]{scipy2020}
Virtanen, P., Gommers, R., Oliphant, T.~E., {et~al.} 2020, Nature Methods, 17,
  261, \dodoi{10.1038/s41592-019-0686-2}

\bibitem[{{Wallerstein}(1978)}]{Wallerstein1978}
{Wallerstein}, G. 1978, \pasp, 90, 36, \dodoi{10.1086/130274}

\bibitem[{{Wannier} {et~al.}(1990){Wannier}, {Sahai}, {Andersson}, \&
  {Johnson}}]{Wannier1990}
{Wannier}, P.~G., {Sahai}, R., {Andersson}, B.~G., \& {Johnson}, H.~R. 1990,
  \apj, 358, 251, \dodoi{10.1086/168980}

\bibitem[{{Webster} \& {Allen}(1975)}]{Webster1975}
{Webster}, B.~L., \& {Allen}, D.~A. 1975, \mnras, 171, 171,
  \dodoi{10.1093/mnras/171.1.171}

\bibitem[{{Wiegert} {et~al.}(2020){Wiegert}, {Groenewegen}, {Jorissen},
  {Decin}, \& {Danilovich}}]{Wiegert2020}
{Wiegert}, J., {Groenewegen}, M.~A.~T., {Jorissen}, A., {Decin}, L., \&
  {Danilovich}, T. 2020, \aap, 642, A142, \dodoi{10.1051/0004-6361/202038029}

\bibitem[{{Yudin} {et~al.}(1994){Yudin}, {Munari}, {Taranova}, \&
  {Dalmeri}}]{Yudin1994}
{Yudin}, B., {Munari}, U., {Taranova}, O., \& {Dalmeri}, I. 1994, \aaps, 105,
  169

\end{thebibliography}
\bibliographystyle{aasjournal}

\appendix
\section{HM Sge observations}
In Table \ref{table:forcast} we list the details of our FORCAST imaging observations. In Tables \ref{table:uv_spec} and \ref{table:exes} we show the results of our COS (UV) and EXES (IR) spectroscopic analyses, respectively.  In Table \ref{table:observations} we have compiled a list of other spectroscopic and imaging observations relevant to this study including the telescope, filter, wavelength range, and exposure time.

\begin{deluxetable}{ccrr}[h]
\tablewidth{\columnwidth}
\tabletypesize{\small}
\tablecolumns{4}
\tablecaption{FORCAST Imaging Observations of HM Sge. \label{table:forcast}}
\tablehead{
\colhead{Date} &
\colhead{Filter} &
\colhead{$\lambda_{\rm effective}$} &
\colhead{$T_{\rm exp.}$}\\
&&\colhead{($\mu$m)} & \colhead{(s)}}
\startdata
2021 Jun 1 &FOR\_F056 & 5.60 & 29 \\
2021 Jun 8 &FOR\_F064 & 6.35 & 146 \\
2021 Jun 8 &FOR\_F077 & 7.69 & 66 \\
2021 Jun 1 &FOR\_F111 & 11.06 & 111 \\
2021 Jun 8 &FOR\_F197 & 19.36 & 41 \\
2021 Jun 8 &FOR\_F253 & 25.33 & 140 \\
2021 Jun 8 &FOR\_F315 & 31.10 & 50 \\
2021 Jun 8 &FOR\_F336 & 33.52 & 219 \\
2021 Jun 8 &FOR\_F348 & 34.58 & 159 \\
2021 Jun 1 &FOR\_F371 & 36.99 & 100 \\
\enddata
\end{deluxetable}

\begin{deluxetable}{cccc}
\tablewidth{\columnwidth}
\tabletypesize{\small}
\tablecolumns{4}
\tablecaption{UV emission lines in the COS spectra of HM Sge. \label{table:uv_spec}}
\tablehead{
\colhead{Ion}
& \colhead{\llap{Wa}velength (\AA)}
& \colhead{Flux\tablenotemark{a}}
& \colhead{Change from 1989\tablenotemark{b}}
}
\startdata
\cutinhead{FUV G140L/800}
O {\footnotesize VI} & 1032 & 5.4 & \nodata \\
He {\footnotesize II} & 1085 & 2.3 & \nodata \\
Ne {\footnotesize V}] & 1136 & 0.7 & \nodata \\
Ne {\footnotesize V}] & 1145 & 1.6 & \nodata \\
Mg {\footnotesize VI} & 1189, 1191 & 7.0 & \nodata \\
N {\footnotesize V} & 1238, 1242 & 45.2 & $+10\%$ \\
O {\footnotesize V} & 1371 & 2.3 & $=$ \\
O {\footnotesize IV}]+Si IV & 1400 & 27.0 & $=$ \\
N {\footnotesize IV}] & 1485 & 27.3 & $-20\%$ \\
C {\footnotesize IV} & 1548,1551 & 111.0 & $-30\%$ \\
{[}Ne {\footnotesize V}{]} & 1575 & 3.6 & $-28\%$ \\
{[}Ne {\footnotesize IV}{]} & 1601 & 2.2 & $-78\%$ \\
He {\footnotesize II} & 1640 & 122.0 & $+36\%$ \\
O {\footnotesize III}] & 1660 & 4.0 & $=$ \\
O {\footnotesize III}] & 1666 & 11.1 & $=$ \\
N {\footnotesize IV} & 1719 & 1.2 & $=$ \\
N {\footnotesize III}] & 1750 & 15.3 & $-24\%$ \\
Mg {\footnotesize VI} & 1806 & 5.4 & new\tablenotemark{c} \\
Si {\footnotesize II} & 1816 & 1.3 & $-35\%$ \\
Si {\footnotesize III}] & 1892 & 8.3 & $=$ \\
C {\footnotesize III}] & 1909 & 42.0 & $-35\%$ \\
\cutinhead{NUV G230L/2950}
{[}Mg {\footnotesize V}{]} & 2929 & 4.7 & $-53\%$ \\
{[}Ne {\footnotesize V}{]} + Fe II & 2973 & 4.6 & $-23\%$ \\
O {\footnotesize III} & 3023 & 2.0 & $-50\%$ \\
O {\footnotesize III} & 3047 & 4.7 & $-48\%$ \\
O {\footnotesize III} & 3133 & 39.0 & $-34\%$ \\
He {\footnotesize II} + Fe II & 3204 & 14.6 & $-36\%$ \\
\enddata
\tablenotetext{a}{Units of $10^{-13}$\,\fluxu}
\tablenotetext{b}{Change since 1989 November, from Table 2 of \citet{Nussbaumer1990}.}
\tablenotetext{c}{This line was not detected in the 1989 IUE spectrum.}
\end{deluxetable}

\begin{deluxetable*}{rlccccccc} 
\tablewidth{\columnwidth}
\tabletypesize{\small}
\tablecolumns{9} 
\tablecaption{Results of EXES Observations \label{table:exes}}
\tablehead{
\colhead{ } &
\colhead{Date} &
\colhead{$v^{\prime}$\,$-$\,$v^{\prime \prime}$} &
\colhead{\emph{J}$_{Ka,Kc}$$^{\prime}$\,$-$\,\emph{J}$_{Ka,Kc}$$^{\prime \prime}$} &
\colhead{$\lambda_{\rm Observed}$} &
\colhead{$\lambda_{\rm Model}$} &
\colhead{Velocity} &
\colhead{Upper State} &
\colhead{Einstein} \\
 &
 &
 &
 &
{} &
{ } &
{Shift$^{*}$}&
{Energy} &
{Coefficient} \\
 &
 &
 &
 &
{ \footnotesize ($\mu$m)} &
{ \footnotesize ($\mu$m)} &
{ \footnotesize (km\,s$^{-1}$) }&
{ \footnotesize (K)} &
{ \footnotesize (A)}}
\startdata
 \multirow{3}{*}{{\Huge \color{light-gray}{\{}}} &2022 Mar 1 & $\upsilon_{2} = 1-0$ & 8$_{0,8}$$-$7$_{1,7}$ & 5.71009 & 5.71107 & \llap{$-$}31.9 & 3363.1 & 12.4 \\
 &2022 Mar 1 & $\upsilon_{2} = 1-0$ & 8$_{0,8}$$-$7$_{1,7}$ & 5.71069 & 5.71107 & \llap{$-$}0.5 & 3363.1 & 12.4 \\
 &2022 Apr 27 & $\upsilon_{2} = 1-0$ & 8$_{0,8}$$-$7$_{1,7}$ & 5.71056 & 5.71107 & 1.9 & 3363.1 & 12.4 \\
 &2022 Mar 1 & $\upsilon_{2} = 1-0$ & 11$_{3,8}$$-$11$_{2,9}$ & 5.71261 & 5.71306 & \llap{$-$}4.3 & 4950.9 & 3.2 \\
 &2022 Mar 1 & $\upsilon_{2} = 1-0$ & 6$_{4,3}$$-$6$_{3,4}$ & 5.71586 & 5.71623 & \llap{$-$}0.3 & 3450.8 & 2.9 \\
 \multirow{3}{*}{{\Huge \color{light-gray}{\{}}}&2022 Mar 1& $\upsilon_{2} = 1-0$ & 5$_{4,2}$$-$5$_{3,3}$ & 5.71988 & 5.72171 & \llap{$-$}76.7 & 3239.7 & 2.4 \\
 &2022 Mar 1 & $\upsilon_{2} = 1-0$ & 5$_{4,2}$$-$5$_{3,3}$ & 5.72074 & 5.72171 & \llap{$-$}31.5 & 3239.7 & 2.4 \\
 &2022 Mar 1 & $\upsilon_{2} = 1-0$ & 5$_{4,2}$$-$5$_{3,3}$ & 5.72131 & 5.72171 & \llap{$-$}1.9 & 3239.7 & 2.4 \\
 \multirow{2}{*}{{\Large \color{light-gray}{\{}}}&2022 Mar 1 & $\upsilon_{2} = 1-0$ & 4$_{4,1}$$-$4$_{3,2}$ & 5.72342 & 5.72383 & \llap{$-$}2.0 & 3064.0 & 1.6 \\
 &2022 Mar 1 & $\upsilon_{2} = 1-0$ & 4$_{4,1}$$-$4$_{3,2}$ & 5.72289 & 5.72383 & \llap{$-$}29.8 & 3064.0 & 1.6 \\
 &2022 Mar 1 & $\upsilon_{2} = 1-0$ & 8$_{1,7}$$-$8$_{0,8}$ & 5.72601 & 5.72643 & \llap{$-$}2.3 & 3583.1 & 1.4 \\
 &2022 Mar 1 & $\upsilon_{2} = 1-0$ & 4$_{4,0}$$-$4$_{3,1}$ & 5.72770 & 5.72811 & \llap{$-$}1.9 & 3064.1 & 1.6 \\
 &2022 Apr 27 & $\upsilon_{2} = 1-0$ & 4$_{4,0}$$-$4$_{3,1}$ & 5.72758 & 5.72811 & 0.5 & 3064.1 & 1.6 \\
 &2022 Mar 1 & $\upsilon_{2} = 1-0$ & 8$_{3,6}$$-$8$_{2,7}$ & 5.73173 & 5.73199 & 5.4 & 3784.3 & 2.7 \\
 \multirow{2}{*}{{\Large \color{light-gray}{\{}}} &2022 Apr 27 & $\upsilon_{2} = 1-0$ & 10$_{3,8}$$-$10$_{4,7}$ & 6.70462 & 6.70516 & 4.2 & 4421.0 & 9.6 \\
 & 2021 Dec 4 & $\upsilon_{2} = 1-0$ & 10$_{3,8}$$-$10$_{4,7}$ & 6.70549 & 6.70516 & 1.9 & 4421.0 & 9.6 \\
 & 2022 Apr 27 & $\upsilon_{2} = 1-0$ & 4$_{0,4}$$-$5$_{1,5}$ & 6.70715 & 6.70770 & 4.1 & 2614.9 & 8.6 \\
 & 2022 Apr 27 & $\upsilon_{2} = 1-0$ & 6$_{1,6}$$-$6$_{2,5}$ & 6.71168 & 6.71212 & 8.3 & 2939.1 & 4.9 \\
 \multirow{2}{*}{{\Large \color{light-gray}{\{}}}& 2022 Apr 27 & $\upsilon_{2} = 1-0$ & 6$_{3,4}$$-$7$_{2,5}$ & 6.71399 & 6.71455 & 3.2 & 3268.5 & 1.9 \\
 & 2021 Dec 4 & $\upsilon_{2} = 1-0$ & 6$_{3,4}$$-$7$_{2,5}$ & 6.71491 & 6.71455 & 3.0 & 3268.5 & 1.9 \\
 \multirow{2}{*}{{\Large \color{light-gray}{\{}}} & 2022 Apr 27 & $\upsilon_{2} = 1-0$ & 8$_{2,7}$$-$8$_{3,6}$ & 6.71519 & 6.71569 & 6.1 & 3590.0 & 7.7 \\
 & 2021 Dec 4 & $\upsilon_{2} = 1-0$ & 8$_{2,7}$$-$8$_{3,6}$ & 6.71611 & 6.71569 & 5.6 & 3590.0 & 7.7 \\
 & 2022 Apr 27 & $\upsilon_{2} = 1-0$ & 2$_{1,1}$$-$3$_{2,2}$ & 6.72284 & 6.72337 & 4.3 & 2436.8 & 8.4 \\
 \multirow{2}{*}{{\Large \color{light-gray}{\{}}} & 2022 Apr 27 & $\upsilon_{2} = 2-1$ & 7$_{1,6}$$-$7$_{2,5}$ &  6.72548 & 6.72611 & 0.2 & 5581.6 & 20.3 \\
 & 2021 Dec 4 & $\upsilon_{2} = 2-1$ & 7$_{1,6}$$-$7$_{2,5}$ & 6.72641 & 6.72611 & 0.1 & 5581.6 & 20.3 \\
 \multirow{2}{*}{{\Large \color{light-gray}{\{}}} & 2022 Apr 27 & $\upsilon_{2} = 1-0$ & 9$_{1,8}$$-$9$_{2,7}$ & 6.72820 & 6.72876 & 3.2 & 3867.6 & 7.9 \\
 & 2021 Dec 4 & $\upsilon_{2} = 1-0$ & 9$_{1,8}$$-$9$_{2,7}$ & 6.72913 & 6.72876 & 3.2 & 3867.6 & 7.9 \\
 & 2021 Dec 4 & $\upsilon_{2} = 2-1$ & 3$_{2,1}$$-$3$_{3,0}$ & 6.73372 & 6.73340 & 1.3 & 4881.5 & 9.1 \\
\multirow{2}{*}{{\Large \color{light-gray}{\{}}} & 2022 Apr 27 & $\upsilon_{2} = 1-0$ & 8$_{4,5}$$-$9$_{3,6}$ & 6.74803 & 6.74865 & 1.0 & 3977.8 & 1.2 \\
 & 2021 Dec 4 & $\upsilon_{2} = 1-0$ & 8$_{4,5}$$-$9$_{3,6}$ & 6.74898 & 6.74865 & 1.4 & 3977.8 & 1.2 \\
 \multirow{2}{*}{{\Large \color{light-gray}{\{}}}& 2022 Apr 27 & $\upsilon_{2} = 1-0$ & 5$_{2,4}$$-$6$_{1,5}$ & 6.75049 & 6.75107 & 2.6 & 2912.3 & 4.5 \\
 & 2021 Dec 4 & $\upsilon_{2} = 1-0$ & 5$_{2,4}$$-$6$_{1,5}$ & 6.75161 & 6.75107 & 10.8 & 2912.3 & 4.5 \\
 & 2022 Mar 5 & $\upsilon_{2} = 1-0$ & 9$_{3,6}$$-$10$_{4,7}$ & 7.55620 & 7.55667 & 2.0 & 4179.2 & 3.9 \\
 & 2022 Mar 5 & $\upsilon_{2} = 1-0$ & 7$_{7,1}$$-$8$_{8,0}$ & 7.57018 & 7.57078 & \llap{$-$}3.0 & 4474.5 & 15.6 \\
 & 2022 Mar 5 & $\upsilon_{2} = 1-0$ & 5$_{1,5}$$-$6$_{2,4}$ & 7.57490 & 7.57543 & \llap{$-$}0.6 & 2766.5 & 0.7 \\
 & 2022 Mar 5 & $\upsilon_{2} = 1-0$ & 10$_{3,7}$$-$11$_{4,8}$ & 7.57958 & 7.58018 & \llap{$-$}3.2 & 4549.8 & 3.7 \\
 & 2022 Mar 5 & $\upsilon_{2} = 1-0$ & 6$_{2,5}$$-$7$_{3,4}$ & 7.58140 & 7.58190 & 0.9 & 3109.6 & 2.2 \\
 & 2022 Mar 5 & $\upsilon_{2} = 1-0$ & 13$_{0,13}$$-$14$_{1,14}$ & 7.58886 & 7.58949 & \llap{$-$}4.3 & 4879.1 & 5.7 \\
 & 2022 Mar 5 & $\upsilon_{2} = 1-0$ & 7$_{4,3}$$-$8$_{5,4}$ & 7.59257 & 7.59318 & \llap{$-$}3.7 & 3700.8 & 7.5 \\
 & 2022 Mar 5 & $\upsilon_{2} = 1-0$ & 7$_{5,2}$$-$8$_{6,3}$ & 7.61816 & 7.61872 & \llap{$-$}1.4 & 3919.5 & 10.1 \\
 & 2022 Mar 5 & $\upsilon_{2} = 1-0$ & 7$_{5,3}$$-$8$_{6,2}$ & 7.61910 & 7.61963 & \llap{$-$}0.0 & 3919.3 & 10.1 \\
\enddata
\tablecomments{Upper state energy is calculated by adding the lower state energy from the HITRAN database to the wavenumber of the feature. The Einstein coefficient is from the listed HITRAN water transition. Curly brackets indicate the same transition. }
\tablenotetext{}{$^{*}$Velocity shifts are corrected for the motion of Earth and the systemic velocity \citep[5\,km\,s$^{-1}$;][]{Hinkle2013}. While additional features may be visible, we focus on those most likely to be rovibrational water transitions.\\}
\end{deluxetable*}
\newcommand{\Mueller}{a}
\newcommand{\Hack}{b}
\newcommand{\thiswork}{c}
\newcommand{\Solf}{d}
\newcommand{\Corradi}{e}
\newcommand{\Schmid}{f}
\newcommand{\Eyres}{g}
\newcommand{\Neugebauer}{h}
\newcommand{\Schild}{i}
\newcommand{\Sacuto}{j}
\newcommand{\Richards}{k}
\newcommand{\Cho}{l}
\newcommand{\Munari}{m}
\newcommand{\Yudin}{n}
\newcommand{\Taranova}{o}

\begin{deluxetable*}{lccrcrrrrr}[h]
\tablewidth{\columnwidth}
\tabletypesize{\ssmall}
\tablecolumns{10}
\tablecaption{Relevant Observations of HM Sge. \label{table:observations}}
\tablehead{
\colhead{\#} &
\colhead{Date} &
\colhead{Telescope} &
\colhead{Instrument} &
\colhead{Type} &
\colhead{Filter} &
\colhead{Wavelength} &
\colhead{Exposure} &
\colhead{Ref.} &
\colhead{Notes} \\
& & & & & & Range & Time (s)
}
\startdata
\textbf{UV} \\
1 & 1978 Jun 6 -- 1983 Oct 15 & IUE & SWP, LWR & Spec. & & 1100--3300\,\AA & & \emph{\Mueller} & \\
2 & 1992 Dec 1 & HST & FOC & Imag. & F190M & 1980\,\AA & 596 & \emph{\Hack} & \\
& & & FOC & Imag. & F253M & 2550\,\AA & 596 & & \\
& & & FOC & Imag. & F195W & 2090\,\AA & & & \\
3 & 2021 Mar 21 & HST & COS & Spec. & G140L & 1230--2050\,\AA & 400\rlap{$^{*}$} & \emph{\thiswork} & \\
 & & & COS & Spec. & G230L & 1700--3200\,\AA & 200\rlap{$^{*}$} & & \\ \hline
\multicolumn{2}{l}{\textbf{Optical}} \\
4 & 1983 Sept 15 & \llap{2.2 m} Calar Alto& Coud\'e & Spec. & \nii, \oiii & 6583, 4959\,\AA & 300--900 & \emph{\Solf}\\
5 & 1996 Jun 4 & NOT & BroCam2 & Imag./Spec. & \nii & 6589\,\AA &90, 300, 900$\times$2 & \emph{\Corradi} & \\
& & & BroCam2 & Imag. & H$\alpha$ cont. & 6507\,\AA & 300 & & \\
6 & 1997 Jul 16 & NOT & ALFOSC & Imag. & [O\,{\sc ii}] & 3725\,\AA & 120, 2400$\times$2 & & \\
& & & ALFOSC & Imag. & [O\,{\sc iii}] & 5012\,\AA & 60, 300 & & \\
& & & ALFOSC & Imag. & [O\,{\sc i}] & 6311\,\AA & 60 & & \\
7 & 1998 Jun 29 -- 1998 Jul 1 & WHT & ISIS & Spec.$^{\dagger}$ & R158R & 6400--9400\,\AA & & \emph{\Schmid} & \\
& & & ISIS & Spec.$^{\dagger}$ & R600R & 6450--7250\,\AA & & & \\
& & & ISIS & Spec.$^{\dagger}$ & R300B & 3300--6300\,\AA & & & \\
8 & 1999 Oct 22 & HST & WFPC2 & Imag. & F218W & 2202\,\AA & 40, 100 & \emph{\Eyres} & (UV) \\
& & & WFPC2 & Imag. & F437W & 4370\,\AA & 800 & & \\
& & & WFPC2 & Imag. & F469W & 4694\,\AA & 100 & & \\
& & & WFPC2 & Imag. & F487W & 4865\,\AA & 200 & & \\
& & & WFPC2 & Imag. & F502W & 5012\,\AA & 100 & & [O\,{\sc iii}] \\
& & & WFPC2 & Imag. & F547W & 5472\,\AA & 2, 20 & & \\
& & & WFPC2 & Imag. & F656N & 6563\,\AA & 100 & & H$\alpha$ \\
9 & 2021 Apr 1 & HST & WFC3 & Imag. & F502N & 5009\,\AA & 80$\times$2 & \emph{\thiswork} & [O\,{\sc iii}] \\
& & & WFC3 & Imag. & F656N & 6561\,\AA & 68 $\times$2 & & H$\alpha$ \\
& & & WFC3 & Imag. & F658N & 6538\,\AA & 60, 3844$\times$3 & & \nii \\ \hline
\textbf{IR} \\
10 & 1983 & IRAS & LRS & Spec. & & 8--23\,$\mu$m & & \emph{\Neugebauer} & \\
11 & 1996 Oct 1 & ISO & SWS & Spec. & & 2.4--45.2\,$\mu$m & 1140 & \emph{\Schild} & \\
& & & LWS & Spec. & & 43.1--195.7\,$\mu$m & 822$\times$2 & & \\
12 & 1997 May 16 & ISO & SWS & Spec. & & 2.4--45.2\,$\mu$m & 1912 & \emph{\Schild} & \\
13 & 2005 Jul 23 -- 2006 Jun 11& VLTI & MIDI & Spec. & \emph{N} band & 5--13\,$\mu$m & & \emph{\Sacuto} & \\
14 & 2021 Jul 1 & SOFIA & FORCAST & Spec. & FOR\_G111 & 8.4--13.7\,$\mu$m & 335 & \emph{\thiswork} & \\
& 2021 Jul 8 & & FORCAST & Spec. & FOR\_G227 & 17.6--27.7\,$\mu$m & 121 & & \\
15 & 2021 Jul 1 -- 2021 Jul 8 & SOFIA & FORCAST & Imag. & All 10 filters & 5.5--38.7\,$\mu$m & 29--220 & \emph{\thiswork} & (Table \ref{table:forcast})\\
16 & 2021 Dec 4 & SOFIA & EXES & Spec. & 8th order & 6.70--6.75\,$\mu$m & 1536 & \emph{\thiswork} & (Table \ref{table:exes})\\
& 2022 Mar 1 & SOFIA & EXES & Spec. & 10th order & 5.70-5.74\,$\mu$m & 1216 & & \\
& 2022 Mar 5 & SOFIA & EXES & Spec. & 7th order & 7.56-7.63\,$\mu$m & 2048 & & \\
& 2022 Apr 27 & SOFIA & EXES & Spec. & 8th order & 6.70-6.76\,$\mu$m & 1024 & & \\
& 2022 Apr 27 & SOFIA & EXES & Spec. & 10th order & 5.70-5.74\,$\mu$m & 320 & & \\ \hline
\multicolumn{2}{l}{\textbf{Radio}} \\
17 & 1992 -- 1997 & \llap{VLA, }MERLIN & & & & \llap{0.4, 1.6, 4.9,} 22.5 GHz & & \emph{\Richards} & \\
18 & 1999 Sep 26 & VLA & & & & \llap{8.56,} 23 GHz & & \emph{\Eyres} & \\
19 & 2009 Dec 05 & KVN Yonsei & & & & \llap{43.1, 42.8} 22.2 GHz & & \emph{\Cho} & \\ \hline
\textbf{M\rlap{onitoring$^{**}$}}\\
20 & 1978 -- 1987 & SAAO & & Imag. & \emph{JHKL} & & & \emph{\Munari} & \\
21 & 1978 -- 1993 & Dalmeri & & Imag. & \emph{\llap{UBVR}JHKL\rlap{MN}} & & & \emph{\Yudin} & \\
22 & 1978 -- 1999 & SAI & & Imag. & \emph{JHKL\rlap{M}} & & & \emph{\Taranova} & \\
\enddata
\tablenotetext{\dagger}{Spectropolarimetry.}
\vspace{-0.2cm}\tablenotetext{*}{Observations included three pointings; exposure times listed are for individual pointing.}
\vspace{-0.2cm}\tablenotetext{$**$}{Data supplemented by the AAVSO.}
\tablerefs{
$^{\Mueller}$\citet{Mueller1985},
$^{\Hack}$\citet{Hack1993},
$^{\thiswork}$\emph{This Work},
$^{\Solf}$\citet{Solf1984},
$^{\Corradi}$\citet{Corradi1999},
$^{\Schmid}$\citet{Schmid2000},
$^{\Eyres}$\citet{Eyres2001},
$^{\Neugebauer}$\citet{Neugebauer1984},
$^{\Schild}$\citet{Schild2001},
$^{\Sacuto}$\citet{Sacuto2007},
$^{\Richards}$\citet{Richards1999},
$^{\Cho}$\citet{Cho2010},
$^{\Munari}$\citet{Munari1989},
$^{\Yudin}$\citet{Yudin1994},
$^{\Taranova}$\citet{Taranova2000}.
}
\end{deluxetable*}
\clearpage


\end{document}